# Wetting and pressure gradient performance in lattice Boltzmann color gradient model

M. Sedahmed*, R. C. V. Coelho[1,2]

*(1) Centro de Física Teórica e Computacional, Faculdade de Ciências, Universidade de Lisboa, 1749-016 Lisboa, Portugal.*
*(2) Departamento de Física, Faculdade de Ciências, Universidade de Lisboa, 1749-016 Lisboa, Portugal.*
**Corresponding author email: mahmoud.sedahmed@alexu.edu.eg*

## ABSTRACT

An accurate implementation of wetting and pressure drop is crucial to correctly reproducing fluid displacement processes in porous media. Although several strategies have been proposed in the literature, a systematic comparison of them is needed to determine the most suitable for practical applications. Here, we carried out numerical simulations to investigate the performance of two widely used wettability schemes in the lattice Boltzmann color gradient model, namely the geometrical wetting scheme by Leclaire et al. (Scheme-I) and the modified direction of the color gradient scheme by Akai et al. (Scheme-II). We showed that Scheme-II was more accurate in simulating static contact angles of a fluid droplet on a solid surface. However, Scheme-I was more accurate in simulating a dynamic case of a binary fluid flow in a horizontal capillary tube described by the Washburn equation. Moreover, we investigated the performance of two popular pressure gradient implementation types. Type-I used the so-called Zou-He pressure boundary conditions at the inlet and the outlet of the domain, while Type-II used an external body force as a pressure gradient. We showed that the Type-I implementation was slightly more accurate in simulating a neutrally wetting fluid in a horizontal capillary tube described by the Washburn equation. We also investigated the differences between the two types of pressure gradient implementation in simulating two fluid displacement processes in a Bentheimer sandstone rock sample: the primary drainage and the imbibition displacement processes.

## I. INTRODUCTION

Multiphase flow through porous medium occurs in industrial and environmental processes like oil and gas production and carbon dioxide capture and storage (CCS), which have a significant economic and environmental impact. Despite the increasing effort to use renewable energies, the world still generates a lot of its energy from oil and gas extracted from underground porous rocks (reservoirs) [1]. Additionally, one of the most promising techniques to reduce global carbon dioxide emissions is collecting it from significant sources and storing it in depleted underground reservoirs. Hence, it is vital to understand the underlying physics involved in these processes. For years, petroleum engineers described the fluid displacement processes in porous media at the macroscopic scale (meters to kilometers) using the well-known Darcy equation. That approach relied on average effective properties such as porosity and permeability. However, much of the underlying physics occurs at the pore scale (micrometers to millimeters), where flow properties on this scale ultimately determine the higher scale properties. Advances in pore-scale physics provide ways to fill the gap between these scales and numerical simulations play an essential role in this context.

Pore scale properties are essential for accurately describing and characterizing fluid flow behavior in reservoirs. Laboratory techniques known as special core analysis (SCAL) offer a detailed description of the reservoir and evaluate meaningful fluid flow relationships such as the capillary pressure–saturation relationship. Special Core Analysis experiments require anywhere between 3 to 18 months to complete and a considerable cost, representing a significant challenge in early production, reservoir development, and predicting reservoir performance [2]. Moreover, the plugging, cleaning, and drying process of the core plugs may change the wettability of core plugs. Wettability expresses the property of one fluid adhering to a rock surface in the presence of another immiscible fluid. Therefore, the wettability type controls the distribution of fluids within the rock pore space.

An essential parameter in reservoir engineering is capillary pressure. Capillary pressure is defined as the pressure difference between the nonwetting phase and the wetting phase as a function of the wetting-phase saturation. In reservoir engineering, capillary pressure is an essential parameter for the study of vertical saturation distribution and all simulation studies. Capillary pressures in a reservoir determine the saturation distribution and the total in-situ volumes of fluids (oil/water/gas). Accurate knowledge of the capillary pressure distribution is one of the primary factors in the reliable estimation of hydrocarbon reserves.

A petroleum reservoir was initially saturated with water. The migration of oil into the reservoir displaced a part of the water. This displacement of the wetting phase (water) by a nonwetting phase (oil) can be simulated in the laboratory experiment of measuring the drainage capillary pressure curve. Drainage is the displacement of a wetting phase by a nonwetting phase (the nonwetting saturation increases). For a water-wet rock, the water saturation decreases. In reservoir studies, the inflow of water must also be modeled. In this case, the imbibition capillary pressure is of interest. Imbibition is the displacement of a nonwetting phase by a wetting phase (the nonwetting saturation decreases). For a water-wet rock, the water saturation increases. Hence, the capillary pressure curve provides vital information in reservoir characterization as it provides the fluid saturation distribution in the reservoir, estimates the largest pores that control permeability, delivers irreducible water saturation and residual oil saturation, and gives information about pore size distribution (sorting).

The increasing complexity of reservoir rocks and the development of tomographic techniques promoted a new type of sample investigation, digital core analysis. High-resolution images of the rock's pores and mineral grains are obtained and processed, and the rock properties are evaluated by numerical simulation at the pore scale. In this work, the color gradient model of the lattice Boltzmann method is used to simulate the displacement process that occurs in petroleum reservoirs and special core analysis laboratory experiments.

The color gradient model (CGM) is one of the most widely used lattice Boltzmann method (LBM) multicomponent flow models, especially for simulating immiscible fluid displacement in porous media. A significant advantage of the color gradient model is that the interfacial tension and contact angle could be set beforehand rather than calibrated with the model parameters, as occurs as in the pseudopotential model [3] [4] [5]. Also, the viscosity ratio could be set independently from other parameters, such as the interfacial tension and contact angle [6]. Moreover, the fluid volumes are well preserved with little mass transfer between the droplets in the CGM, which is required for an accurate description of fluid displacement processes. Such advantages make the CGM a very appealing choice for digital rock models [7] [8].

The first version of the color gradient model was introduced by Gunstensen et al. [9], where two distribution functions (red and blue) were introduced to represent two different fluids, and each fluid went through the collision and streaming steps. The original model suffered from several drawbacks, such as high spurious currents near the interface, lattice pinning, and lack of Galilean invariance [10]. Several improvements were presented in the literature to alleviate many of the drawbacks of the original CGM. For instance, the original perturbation operator was criticized for not recovering the two-phase Navier-Stokes equations [11] [12]. Reis and Philips [11] proposed an improved perturbation operator to obtain the correct interfacial force. Liu et al. [12] used the concept of continuum surface force (CSF) [13] and the constraints of mass and momentum conservation to derive a generalized perturbation operator. Lishchuk et al. [14] also used the concept of the continuum surface force to model the interfacial tension, where the perturbation step was replaced by a direct forcing term in the mixed region. That model was shown to reduce the spurious currents and improve the isotropy of the interface. Tölke et al. [15] proposed an improved recoloring algorithm to minimize the spurious currents in the interface region. Latva-Kokko and Rothman [10] proposed another improved recoloring operator to reduce the lattice pinning phenomena. Leclaire et al. [16] showed that the Latva-Kokko and Rothman [10] recoloring scheme significantly improved the numerical stability and accuracy of the color gradient model, reduced the spurious currents, and eliminated the lattice pinning issue.

Many variants of the color gradient model exist in the literature [15] [17] [18] [19] [8] [20] [7]. We chose the model presented by Chen et al. [20]. They presented a version of the color gradient model where the continuum surface force (CSF) was combined with a geometrical wetting boundary condition [19] under the framework of multiple relaxation time (MRT) collision operator [21] and the Latva-Kokko and Rothman [10] recoloring scheme. The model represents a state-of-the-art color gradient model that is numerically stable, accurate, and well-suited for simulating fluid flow in porous media. An open-source software is available where the model is implemented, known as the microfluid LBM solver (MF-LBM) [20]. It was written in the Fortran 90 language. We developed a newer version of the solver written in the C++ language [22].

In addition to the different collision operators, perturbation operators, and recoloring schemes of the CGM, a wetting boundary condition is needed to simulate the wettability difference of the simulated fluids properly. Hence, the wetting model has a crucial role in the simulations. Several wetting models were introduced in the literature for the CGM. The most popular wetting boundary condition is the fictitious density (FD) model [18] due to its simplicity and ease of implementation. However, it was shown in the literature that this model suffers from a severe lack of accuracy due to a nonphysical fluid transfer along the solid wall [23] [19] [24]. Several geometrical wetting models were introduced in the literature to alleviate the drawbacks of the FD wetting model [23] [19] [24] [25]. We focused on two enhanced wetting models presented by Leclaire et al. [19] and Akai et al. [24]. Those generalized three-dimensional (3D) models were shown to give accurate wettability modeling results. However, they were not directly compared in unified benchmarks to assess their accuracy. In this work, we compared the simulation results of both models in both static and dynamic flow benchmarks.

Inlet and outlet boundary conditions are another important aspect in fluid displacement simulations. For instance, a drainage displacement process in a porous medium occurs due to a pressure gradient along the fluid flow, where a non-wetting fluid displaces a wetting fluid. Here, this pressure gradient was modeled in the LBM via two widely used techniques. In the first one, Dirichlet pressure boundary conditions are applied at the inlet and the outlet of the domain. For instance, the so-called Zou-He technique [26] is often used to implement a constant pressure boundary condition. This technique was used in several studies to implement a constant pressure boundary condition [27] [28] [4]. For example, Huang et al. [27] investigated immiscible fluid displacement in a 2D-generated heterogeneous porous medium, and the flow was driven by applying the Zou-He technique to both the inlet and the outlet boundaries. The second technique for pressure gradient implementation we tested is to apply a uniform external body force on the flow field that provides the effect of the pressure gradient [29]. This technique is more widely used in the literature due to its simplicity and stability [29] [30] [31] [32]. For instance, Ramstad et al. [29] simulated immiscible two-phase flow in 3D digitized rock samples of a Bentheimer sandstone to obtain capillary pressure and relative permeability curves. However, the two techniques were not directly compared in the literature for benchmarks to assess their accuracy. We compared the accuracy of both methods using a benchmark with an analytical solution, which is fluid flow in a horizontal capillary tube under a pressure gradient as described by the Washburn equation [33].

In addition to testing the pressure gradient implementation techniques in known benchmarks, we simulated fluid flow displacement processes in realistic porous media. Two displacement processes were simulated, namely the primary drainage and the imbibition. Drainage is the displacement of a wetting phase by a nonwetting phase (the nonwetting saturation increases), while imbibition is the displacement of a nonwetting phase by a wetting phase (the nonwetting saturation decreases). We used a Bentheimer sandstone rock sub-sample to carry out the simulations. We fixed all simulation parameters and changed only the pressure gradient implementation to illustrate the difference in results due to changing the implementation types at the same pressure gradients. This gives an insight into the impact of the pressure gradient implementation on porous media simulations.

This paper is organized as follows: In Sec. II, we present the color gradient model, the schemes of the wetting models, and the types of pressure gradient implementations. In Sec. III, we provide model validation results using the Laplace law and the static contact angle tests. In Sec. IV, we show simulation results of a binary fluid flow using the different wetting models and pressure gradient implementations in both a horizontal capillary tube and a realistic porous medium. In Sec. V, we give the conclusion of the investigation.

## II. Methodology

### A. Multiple relaxation time Color Gradient LBM model

A three-dimensional color gradient multiphase model (CGM) in the framework of the multiple relaxation time (MRT) collision operator is used in this work, as described in Ref. [20]. In the CGM, different fluids are labeled with different colors, e.g., red fluid (denoted with superscript "r") and blue fluid (denoted with superscript "b"). Each of the fluids is represented by its distribution function $f_i^s$, where "s" represents either fluid, i.e., $f_i^r$ and $f_i^b$, and the subscript "$i$" represents the index of the discrete velocity set. The total distribution function for the fluid mixture is $f_i^t = f_i^r + f_i^b$.

The lattice Boltzmann equation (LBE) used in this work for both fluid "r" and fluid "b" is described as follows

$$f_i^s(\boldsymbol{x}+\boldsymbol{e}_i\,\Delta t, t+\Delta t) = f_i^{s\prime}\big(f_i^{s\prime\prime}(\boldsymbol{x},t)\big), \tag{1}$$

where, $f_i^{s\prime}$ is the fluid distribution function after the CGM recoloring step, $f_i^{s\prime\prime}$ is the fluid post-collision distribution function. The right-hand side of Eq. (1) is evaluated by first determining the post-collision distribution functions $f_i^{s\prime\prime}(\boldsymbol{x},t)$ using the MRT collision operator, which is responsible for the viscous effects. The generation of surface tension is also achieved using the continuum surface force (CSF) model introduced in Ref. [20] in the framework of the MRT collision model. Then, a recoloring step is applied to $f_i^{s\prime\prime}(\boldsymbol{x},t)$ using the recoloring operator demonstrated in Ref. [18] to evaluate $f_i^{s\prime}$. Finally, the distribution functions are propagated to the neighbor lattice points in a streaming step. The model implementation details are shown next.

It should be noted that the MRT collision model shown in Eq. (1) is used to evaluate the total distribution function $f_i^{t\prime\prime}$ instead of the fluid distribution function $f_i^{s\prime\prime}$, as it is assumed in this work that both fluids have equal density. Hence, it is unnecessary to compute the collision operator separately for each fluid component [8] [20]. The MRT collision step is expressed as follows

$$f_i^{t\prime\prime}(\boldsymbol{x},t) = f_i^t(\boldsymbol{x},t) - \boldsymbol{M}^{-1}\boldsymbol{S}\left(\boldsymbol{m}(\boldsymbol{x},t) - \boldsymbol{m}^{eq}(\boldsymbol{x},t) + \Delta t\,\boldsymbol{S}^{\boldsymbol{F}}(\boldsymbol{x},t)\right), \tag{2}$$

where, $\boldsymbol{M}$ is a transformation matrix used to transform $f_i^t$ into the moment space ($\boldsymbol{m} = \boldsymbol{M}\boldsymbol{f}$), and it is given for the used $D_3Q_{19}$ lattice arrangement as shown in Appendix A. The corresponding 19 moments $\boldsymbol{m}$ of the distribution functions are ordered as follows,

$$\boldsymbol{m} = \big(\rho, e, \epsilon, j_x, q_x, j_y, q_y, j_z, q_z, 3p_{xx}, 3\pi_{xx}, p_{ww}, \pi_{ww}, p_{xy}, p_{yz}, p_{xz}, m_x, m_y, m_z\big)^T, \tag{3}$$

where, the zeroth-order moment corresponds to the mass density ($m_0 = \rho$), the three first-order moments correspond to momentum $m_{3,5,7} = j_{x,y,z}$, the six second-order moments $m_1 = e, m_9 = 3p_{xx}, m_{11} = p_{ww} = p_{yy} - p_{zz}$, and $m_{13,14,15} = p_{xy,yz,xz}$ correspond to part of the kinetic energy, the diagonal and off-diagonal elements of the viscous stress tensor, respectively. Moments higher than the second order are called higher-order moments. They do not have a clear physical connection to the incompressible limit of Navier-Stokes equations. However, they are characterized by their dependence on the gradient of the conserved moments. These higher-order moments are the six third-order moments $m_{4,6,8} = q_{x,y,z}, m_{16,17,18} = m_{x,y,z}$ and three fourth-order moments $m_2 = \epsilon, m_{10} = 3\pi_{xx}, m_{12} = \pi_{ww}$ [34]. $\boldsymbol{j} = \rho_o\big(u_x, u_y, u_Z\big)$ is the momentum, where $\rho_o$ is a constant reference density. The density fluctuation $\rho$ is decoupled from the momentum $\boldsymbol{j}$, which is similar to an incompressible LB model with a modified equilibrium distribution function ($\boldsymbol{f}^{eq}$) [35] [36]. The equilibrium moments $\boldsymbol{m}^{eq}$ are evaluated by transforming the equilibrium distribution functions ($\boldsymbol{f}^{eq}$) to the moment space ($\boldsymbol{m}^{eq} = \boldsymbol{M}\boldsymbol{f}^{eq}$). The details of evaluating $\boldsymbol{m}^{eq}$ are shown in Appendix A. $\boldsymbol{S}$ is a diagonal relaxation matrix where each moment is relaxed using an individual relaxation rate as follows

$$\boldsymbol{S} = diag\big(0, S_e, S_\epsilon, 0, S_q, 0, S_q, 0, S_q, S_\nu, S_\pi, S_\nu, S_\pi, S_\nu, S_\nu, S_\nu, S_m, S_m, S_m\big). \tag{4}$$

The forcing term $\boldsymbol{S^F}$ is expressed as shown in Appendix A using an external body force $\boldsymbol{F}$. This force is evaluated using the continuum surface force (CSF) model shown in Ref. [20], where Guo's forcing scheme [37] is employed to implement the CSF in the MRT framework. The model reduces the spurious currents and improves the interface's isotropy [8] [20]. In the CSF model, an external body force is added to the LBE to create the local stress jump across the interface, which can be written as

$$\boldsymbol{F}(\boldsymbol{x}) = \frac{1}{2}\gamma\,\kappa\,\boldsymbol{C}(\boldsymbol{x}), \tag{5}$$

where, $\gamma$ is the surface tension, $\kappa$ is the interface curvature, and $\boldsymbol{C}$ is the color gradient. The color gradient is evaluated as follows

$$\boldsymbol{C}(\boldsymbol{x}) = \nabla\phi(\boldsymbol{x}) = \frac{3}{c^2\Delta t}\sum_i w_i\,\boldsymbol{e}_i\,\phi(\boldsymbol{x} + \boldsymbol{e}_i\,\Delta t), \tag{6}$$

where, $c = \frac{\Delta x}{\Delta t}$ is the lattice speed, $\boldsymbol{e}_i$ represents the lattice velocity vectors, $w_i$ represents the lattice weights, and $\phi$ is the phase field order parameter. The details of $\boldsymbol{e}_i$ and $w_i$ for the used $D_3Q_{19}$ lattice arrangement are shown in Appendix A. The phase field order parameter $\phi$ is determined by

$$\phi(\boldsymbol{x}) = \frac{\rho^r(\boldsymbol{x}) - \rho^b(\boldsymbol{x})}{\rho(\boldsymbol{x})}, \tag{7}$$

which varies between 1 for a pure red "r" fluid and -1 for a pure blue "b" fluid. The fluid component density and the bulk density are given, respectively, by

$$\rho^s(\boldsymbol{x}) = \sum_i f_i^s(\boldsymbol{x}), \tag{8}$$

$$\rho(\boldsymbol{x}) = \rho^r(\boldsymbol{x}) + \rho^b(\boldsymbol{x}). \tag{9}$$

The interface curvature $\kappa$ is defined as follows

$$\kappa = \left[\left(\boldsymbol{I} - \boldsymbol{nn}(\boldsymbol{x})\right)\cdot\nabla\right]\cdot\boldsymbol{n}(\boldsymbol{x}), \tag{10}$$

where, $\boldsymbol{n}$ is the normal direction vector of the fluid interface, and it is defined as

$$\boldsymbol{n}(\boldsymbol{x}) = \frac{\nabla\phi(\boldsymbol{x})}{|\nabla\phi(\boldsymbol{x})|}. \tag{11}$$

By expressing Eq. (10) in three dimensions, we obtain the following expression

$$\begin{aligned}\kappa(\boldsymbol{x}) = {} & (n_x^2(\boldsymbol{x}) - 1)\frac{\partial}{\partial x} n_x(\boldsymbol{x}) + (n_y^2(\boldsymbol{x}) - 1)\frac{\partial}{\partial y} n_y(\boldsymbol{x}) + (n_z^2(\boldsymbol{x}) - 1)\frac{\partial}{\partial z} n_z(\boldsymbol{x}) \\ & + \left(\frac{\partial}{\partial y} n_x(\boldsymbol{x}) + \frac{\partial}{\partial x} n_y(\boldsymbol{x})\right) n_x(\boldsymbol{x}) n_y(\boldsymbol{x}) \\ & + \left(\frac{\partial}{\partial z} n_x(\boldsymbol{x}) + \frac{\partial}{\partial x} n_z(\boldsymbol{x})\right) n_x(\boldsymbol{x}) n_z(\boldsymbol{x}) \\ & + \left(\frac{\partial}{\partial y} n_z(\boldsymbol{x}) + \frac{\partial}{\partial z} n_y(\boldsymbol{x})\right) n_z(\boldsymbol{x}) n_y(\boldsymbol{x}).\end{aligned} \quad (12)$$

It should be noted that the bulk fluid mixture viscosity is related to one of the MRT relaxation rates as follows

$$\frac{1}{S_\nu(\boldsymbol{x})} = \frac{3\,\nu(\boldsymbol{x})}{c^2 \Delta t} + \frac{1}{2}, \quad (13)$$

where a harmonic mean is used to determine the viscosity of the bulk fluid mixture ($\nu$) and account for unequal viscosities of the two fluids [38] as follows

$$\frac{1}{\nu(\boldsymbol{x})} = \frac{1 + \phi(\boldsymbol{x})}{2\,\nu^r} + \frac{1 - \phi(\boldsymbol{x})}{2\,\nu^b}. \quad (14)$$

Other relaxation rates can be tuned to improve accuracy and stability. In this work, we use the values used in Refs. [39] [40] [34] as follows

$$S_e(\boldsymbol{x}) = S_\epsilon(\boldsymbol{x}) = S_\pi(\boldsymbol{x}) = S_\nu(\boldsymbol{x}), \quad (15)$$

$$S_q(\boldsymbol{x}) = S_m(\boldsymbol{x}) = 8\frac{(2 - S_\nu(\boldsymbol{x}))}{(8 - S_\nu(\boldsymbol{x}))}. \quad (16)$$

After evaluating $f_i^{t\prime\prime}$ using the MRT collision model, a recoloring step is applied using the recoloring operator demonstrated in Ref. [18] to evaluate $f_i^{s\prime}$ as follows

$$f_i^{r\prime}(\boldsymbol{x}) = \frac{\rho^r(\boldsymbol{x})}{\rho(\boldsymbol{x})} f_i^{t\prime\prime}(\boldsymbol{x}) + \beta\, w_i \frac{\rho^r(\boldsymbol{x})\rho^b(\boldsymbol{x})}{\rho(\boldsymbol{x})} \cos(\varphi)(\boldsymbol{x})\,|\boldsymbol{e}_i|, \quad (17)$$

$$f_i^{b\prime}(\boldsymbol{x}) = \frac{\rho^b(\boldsymbol{x})}{\rho(\boldsymbol{x})} f_i^{t\prime\prime}(\boldsymbol{x}) - \beta\, w_i \frac{\rho^r(\boldsymbol{x})\rho^b(\boldsymbol{x})}{\rho(\boldsymbol{x})} \cos(\varphi)(\boldsymbol{x})\,|\boldsymbol{e}_i|, \quad (18)$$

where, $\beta$ is the segregation parameter related to the interface thickness, it can take values between 0 and 1. In the simulation cases in this work, $\beta$ was set to 0.95 [20] to generate a relatively sharp fluid interface. $\varphi$ is the angle between the color gradient and the lattice vector, and it can be determined as follows

$$\cos(\varphi)(\boldsymbol{x}) = \frac{\boldsymbol{e}_i \cdot \nabla\phi(\boldsymbol{x})}{|\boldsymbol{e}_i||\nabla\phi(\boldsymbol{x})|}. \quad (19)$$

Substituting Eq. (19) in Eqs. (17) and (18) and considering Eqs. (11) and (9), the following simplified expressions could be obtained by summing Eqs. (17) and (18)

$$f_i^{r\prime}(\boldsymbol{x}) = \frac{\rho^r(\boldsymbol{x})}{\rho(\boldsymbol{x})} f_i^{t\prime\prime}(\boldsymbol{x}) + \beta\, w_i \frac{\rho^r(\boldsymbol{x})\rho^b(\boldsymbol{x})}{\rho(\boldsymbol{x})}\big(\boldsymbol{e}_i \cdot \boldsymbol{n}(\boldsymbol{x})\big), \tag{20}$$

$$f_i^{b\prime}(\boldsymbol{x}) = f_i^{t\prime\prime}(\boldsymbol{x}) - f_i^{r\prime}(\boldsymbol{x}). \tag{21}$$

The above expressions are valid for $i \neq 0$. For $i = 0$, the second term on the right-hand side of the previous expressions vanishes, and only the first term becomes relevant. This is vital for proper coding of the model and for faster computation.

The fluid velocity vector ($\boldsymbol{u}$) is determined by

$$\boldsymbol{u}(\boldsymbol{x}) = \frac{1}{\rho(\boldsymbol{x})}\sum_i \boldsymbol{e}_i f_i^t\,(\boldsymbol{x}) + \frac{\Delta t}{2\,\rho(\boldsymbol{x})}\boldsymbol{F}(\boldsymbol{x}), \tag{22}$$

and the pressure is evaluated as follows

$$P(\boldsymbol{x}) = \frac{c^2\rho(\boldsymbol{x})}{3}. \tag{23}$$

### B. Wetting models

Despite the bounce-back boundary condition is implemented at solid walls to mimic the no-slip boundary condition at walls [41]. It is necessary to add a wetting model to mimic the wettability preference between the different fluids and the solid walls. One of the CGM's most widely used wetting models is the fictitious-density (FD) model due to its simplicity [18]. In the FD model, a specific value of the phase field order parameter $\phi$ is assigned to the solid boundary points as follows

$$\phi(\boldsymbol{x}_s) = \cos(\theta), \tag{24}$$

where, $\boldsymbol{x}_s$ represents the solid boundary points and $\theta$ is the contact angle. However, it was shown in the literature that this model suffers from a severe lack of accuracy due to a nonphysical fluid transfer along the solid wall [23] [19] [24]. Such a deficiency can severely harm the simulation results, especially during simulating fluid displacement in porous media.

Several geometrical wetting models were introduced in the literature to alleviate the drawbacks of the FD wetting model [23] [19] [24] [25]. In this work, we used two enhanced wetting models that we called Scheme-I, presented by Leclaire et al. in Ref. [19], and Scheme-II, presented by Akai et al. in Ref. [24]. Moreover, it was shown in Ref. [20] that two vital steps are needed when combining the presented CGM with any of the wetting models. We will describe the implementation details in the following sections.

#### 1. Scheme-I: Geometrical wetting scheme by Leclaire et al. 2017

This wetting model was introduced by Leclaire et al. in Ref. [19]. The basic idea of this model is to alter the unit normal vector to the fluid interface numerically $(\boldsymbol{n}_c)$ near the solid boundary to set the desired contact angle. Hence, the contact angle is set by altering the orientation of the color gradient $\boldsymbol{C} = |\boldsymbol{C}|\, \boldsymbol{n}_c$ so that its angle with the normal direction to the solid wall $\boldsymbol{n}_w$ is equal to the desired contact angle. This approach could be considered a sort of Dirichlet boundary condition for the orientation of the color gradient, and it should be noted that the norm of the color gradient $|\boldsymbol{C}|$ remains unchanged by the wetting boundary condition. The implementation details of this model are explained in the following paragraphs.

Firstly, the normal vector to the solid wall $\boldsymbol{n}_w$ is estimated for the fluid-solid binary matrix representing the geometry. Such a binary matrix is obtained easily for simple geometries such as a pipe or from processing the Micro-Computed Tomography (μCT) scans of more complicated geometries such as a porous medium. To estimate $\boldsymbol{n}_w$ at the fluid lattice points near the solid boundary, the three-dimensional solid matrix image is first smoothed using the following smoothing operator

$$\boldsymbol{g}(x,y,z)^{(itr)} = \sum_{k=-1}^{k=1}\sum_{j=-1}^{j=1}\sum_{i=-1}^{i=1} w_s(i^2+j^2+k^2)\ g(x+i,y+j,z+k)^{(itr-1)}, \qquad (25)$$

where, $w_s(0) = \frac{8}{27}$, $w_s(1) = \frac{2}{27}$, $w_s(2) = \frac{1}{54}$, $w_s(3) = \frac{1}{216}$. $\boldsymbol{g}(x,y,z)$ is the fluid-solid binary matrix, and $itr$ is the smoothing iteration number. The solid geometry is smoothed for three iterations in this work, as in Ref. [19]. Then the normal $\boldsymbol{n}_w$ to the solid matrix at the fluid lattice points near the boundary are estimated as the gradient of the smoothed image as follows

$$\boldsymbol{n}_w(x,y,z) = \boldsymbol{\nabla} g(x,y,z)^3. \qquad (26)$$

This procedure to estimate $\boldsymbol{n}_w$ is carried out once during the initial step of the computations before the main time loop. Then, the normal unit vector to the fluid interface $(\boldsymbol{n}_c)$ is altered at each time step using the following procedure. Firstly, the current values of $\boldsymbol{n}_c$ is estimated at a fluid point near the boundary as

$$\boldsymbol{n}_c = \frac{\boldsymbol{C}}{|\boldsymbol{C}|}. \qquad (27)$$

Next, the first step of the algorithm is to find a new vector $\boldsymbol{v}_c$ that replaces the estimation of $\boldsymbol{n}_c$ from Eq. (27). The new vector needs to respect the following equation

$$f(\boldsymbol{v}_c) = \boldsymbol{v}_c \cdot \boldsymbol{n}_w - |\boldsymbol{v}_c|\cos(\theta) = 0. \qquad (28)$$

One method to solve Eq. (28) is using the numerical method of secant using the recurrence relation shown in Ref. [19] as a function of the index $n$ as follows

$$\boldsymbol{v}_c^{(0)} = \boldsymbol{n}_c, \qquad (29)$$

$$\boldsymbol{v}_c^{(1)} = \boldsymbol{n}_c - \lambda(\boldsymbol{n}_c + \boldsymbol{n}_w), \qquad (30)$$

$$\boldsymbol{v}_c^{(n)} = \frac{\boldsymbol{v}_c^{(n-2)} f\left(\boldsymbol{v}_c^{(n-1)}\right) - \boldsymbol{v}_c^{(n-1)} f\left(\boldsymbol{v}_c^{(n-2)}\right)}{f\left(\boldsymbol{v}_c^{(n-1)}\right) - f\left(\boldsymbol{v}_c^{(n-2)}\right)}, \qquad (31)$$

where, $\lambda = \frac{1}{2}$ as used in Ref. [19], and the algorithm stops at n = 4. Finally, the vector $\boldsymbol{v}_c^{(4)}$is normalized before it replaces the current orientation of the color gradient $\boldsymbol{n}_c$. This numerical procedure is executed at each time step after approximating the color gradient $\boldsymbol{C}$. Careful programming of the model is needed to avoid undefined behavior of the code in cases where the normal direction of the solid surface is aligned with the fluid interface direction or the evaluated fluid interface direction is already in the required direction.

**2. Scheme-II: Modified direction of the color gradient by Akai et al. 2018**

This wetting model was introduced by Akai et al. [24]. The basic idea of this model is like scheme-I, which is altering the unit normal vector to the fluid interface ($\boldsymbol{n}_c$) near the solid boundary. However, this scheme uses a more simplified technique to estimate $\boldsymbol{n}_c$.
Like scheme-I, the model starts with determining the normal vector to the solid wall $\boldsymbol{n}_w$ as shown in Eqs. (25) – (26). Then, the normal unit vector to the fluid interface ($\boldsymbol{n}_c$) is initially estimated via Eq. (27). Next, two unit vectors $\boldsymbol{n}_\pm$ are evaluated as follows

$$\boldsymbol{n}_\pm = \left(\cos(\pm\theta) - \frac{\sin(\pm\theta)\cos(\theta')}{\sin(\theta')}\right)\boldsymbol{n}_w + \frac{\sin(\pm\theta)}{\sin(\theta')}\boldsymbol{n}_c\,, \tag{32}$$

$$\theta' = \arccos(\boldsymbol{n}_w \cdot \boldsymbol{n}_c). \tag{33}$$

Finally, the Euclidean distances between $\boldsymbol{n}_\pm$ and $\boldsymbol{n}_c$ are evaluated, and $\boldsymbol{n}_c$ is replaced by either $\boldsymbol{n}_+$ or $\boldsymbol{n}_-$, whichever has the shorter Euclidean distance to $\boldsymbol{n}_c$. It could be noted that this scheme requires fewer computations than the secant iterative method presented in scheme-I.

**3. Additional steps for wettability modeling enhancements**

It was shown in Ref. [20] that additional steps are vital in enhancing the wettability models for the CGM. The first step is extrapolating the value of the order parameter $\phi$ on the solid boundary from the nearby fluid boundary points instead of having an assigned value as in the FD wetting model. That would ensure no strong interactions between the fluid and solid points. The value of the order parameter is obtained via weighted sum as follows

$$\phi(\boldsymbol{x}_s) = \frac{\sum_{i:x+\boldsymbol{e}_i\Delta t\,\in\,x_F} w_i\,\phi(x+\boldsymbol{e}_i\Delta t)}{\sum_{i:x+\boldsymbol{e}_i\Delta t\,\in\,x_f} w_i}\,, \tag{34}$$

where, $\boldsymbol{x}_F$ is a fluid boundary point that is in contact with the solid boundary point $\boldsymbol{x}_s$. The summation is carried out to include the phase order parameter values from only neighbor fluid points next to the local solid point.

The second step is applying the zero-interfacial-force scheme shown in Ref. [42] to reduce the unbalanced forces near the three-phase contact line and minimize the spurious currents. That is essential for geometrical wetting models at contact angles far from 90°. This step is carried out by extrapolating the normal direction vector $\boldsymbol{n}_c$ from the nearby fluid points to the solid point as follows

$$\boldsymbol{n}_c(\boldsymbol{x}_s) = \frac{\sum_{i:x+\boldsymbol{e}_i\Delta t\,\in\,x_F} w_i\,\boldsymbol{n}_c(\boldsymbol{x}+\boldsymbol{e}_i\Delta t)}{\sum_{i:x+\boldsymbol{e}_i\Delta t\,\in\,x_f} w_i}. \tag{35}$$

To summarize, the implementation of the above wetting models is applied as follows

1- Extrapolate the phase field order parameter $\phi$ to solid boundary points using Eq. (34).

2- Calculate the normal unit vector to the fluid interface $(\boldsymbol{n}_c)$ using Eq. (27).
3- Alter the normal unit vector to the fluid interface $(\boldsymbol{n}_c)$ in fluid boundary points near the solid boundary points to set the desired contact angle using Eqs. (28) – (31) for scheme-I and Eqs. (32) – (33) for scheme-II.
4- Extrapolate the normal direction vector $\boldsymbol{n}_c$ from the nearby fluid boundary points to the solid boundary points using Eq. (35).
5- Determine the interface curvature $\kappa$ via Eq. (12).

**C. Pressure gradient implementation**

In this work, pressure-driven flow can be simulated using two widely used types of pressure gradients. In the following sections, we describe the implementation of both types of pressure gradients.

**1. Type-I: Pressure boundary conditions**

Dirichlet pressure boundary conditions are used at the inlet and outlet boundary conditions of this type of pressure gradient implementation. We use the term pressure gradient in this type, but practically, this implementation simulates a pressure drop along the flow. For instance, an outlet pressure is fixed to a reference value, while the inlet pressure is increased above this reference value to implement a pressure gradient for the flow. Translating the macroscopic pressure values at the boundaries to distribution functions at the LBE could be achieved via several techniques. In this work, we use the non-equilibrium bounce-back method (NEBB). It is one of the most widely used techniques to apply Dirichlet boundary conditions in LB simulations. It was introduced by Zou and He [26] and is known in the literature as the Zou-He method. The implementation details of the method for the $D_3Q_{19}$ lattice arrangement could be found in Refs. [43] [20]. A summary of the used equations to implement inlet and outlet boundary conditions is shown in Appendix B. It should be noted that implementing a pressure gradient using this implementation results in a density gradient in the fluid, which may not be consistent with the incompressible fluid assumption.

**2. Type-II: External body forces**

In this implementation, the pressure gradient is applied using an external body force exerted on the fluid without having an explicit pressure gradient at the inlet and outlet boundaries. The exerted body force could be evaluated as

$$\boldsymbol{g} = \frac{\nabla P}{\rho^F} = \frac{\Delta P}{L\,\rho^F}, \tag{36}$$

where, $\Delta P$ is the pressure difference between the inlet and the outlet boundaries, $L$ is the distance between the inlet and outlet boundaries, and $\rho^F$ is the fluid density. In this work, the external body force is applied uniformly on both fluids at all fluid lattice points, and it is added to the force term in Eq. (22).

**D. Additional boundary conditions**

**1. Injected fluid at the inlet boundary**

In the simulations carried out in this work, the fluid component injected at the inlet boundary is specified using the phase order parameter $\phi(\boldsymbol{x}_{in})$ and the fraction $\zeta^s$ shown in Eq. (B.1). For instance, injecting a pure fluid component "r" at the inlet boundary is achieved by setting $\phi(\boldsymbol{x}_{in}) = 1$, $\zeta^r = 1$ and $\zeta^b = 0$. Moreover, the injected fluid at the inlet boundary could be replaced in a restart simulation. For instance, it could be needed to simulate injecting sequential patches of different fluid components. This could be achieved as follows

$$f_i^s(\boldsymbol{x}_{in}) = f_i^s(\boldsymbol{x}_{in}) + f_i^{\bar{s}}(\boldsymbol{x}_{in}), \quad (37)$$

$$f_i^{\bar{s}}(\boldsymbol{x}_{in}) = 0, \quad (38)$$

where, $f_i^s(\boldsymbol{x}_{in})$ are the distribution functions of the newly injected fluid component " $s$ ", and $f_i^{\bar{s}}(\boldsymbol{x}_{in})$ are the distribution functions of the previously injected fluid component " $\bar{s}$ ".

**2. Zero gradient phase order parameter at the outlet boundary**

In this work, the phase order parameter $\phi$ at the outlet boundary is set via a simple zero gradient rule as follows

$$\phi(\boldsymbol{x}_{out}) = \phi(\boldsymbol{x}_{out} - 1). \quad (39)$$

This allows the fluid interface to exit the outlet boundary without obstruction.

**E. Physical Units conversion**

Since LBM works on the mesoscopic scale, it also provides all results on that scale. A conversion would be needed from the LBM units to physical units. This is an essential step in comparing simulation results with real-world parameters. In this work, the conversion approach shown in Ref. [29] is adopted, where results are scaled via length unit $a_o$, mass unit $m_o$ and time unit $t_o$. These units are defined as follows

$$a_o = \frac{l_{phys}}{x_{lbm}}, \quad (40)$$

$$m_o = \frac{\rho_{phys}\, a_o^3}{\rho_{lbm}}, \quad (41)$$

$$t_o = \frac{\nu_{lbm}\, a_o^2}{\nu_{phys}}, \quad (42)$$

where, $x_{lbm}$ is the length on the LBM scale, it is measured in lattice length unit $(l.u)$, and it is usually set to unity, $l_{phys}$ is the physical characteristic length, and it is measured in meters $(m)$, $\rho_{lbm}$ is the lattice fluid density, it is measured in lattice mass unit per cubic lattice length unit $(m.u./l.u.^3)$ and is usually set to unity, $\rho_{phys}$ is the physical fluid density, and it is measured in kilograms per cubic meters $(kg/m^3)$, and it is assumed in this work as $\left(\rho_{phys}^r = \rho_{phys}^b = 1000\ kg/m^3\right)$, $\nu_{lbm}$ is the lattice fluid kinematic viscosity and it is measured in squared lattice length unit per lattice time unit $(l.u^2/t.u)$, and it is evaluated using Eq. (13), and $\nu_{phys}$ is the fluid physical kinematic viscosity, and it is measured in squared meters per second $(m^2/s)$. The physical pressure could be obtained as

$$P_{phys} = \frac{P_{lbm}\, m_o}{a_o\, t_o^2}, \tag{43}$$

where, $P_{lbm}$ is the pressure value on the lattice scale, and it is measured in $(m.u/l.u\ t.u.^2)$, and $P_{phys}$ is the physical pressure value, and it is measured in Pascals $(Pa)$. Moreover, interfacial tension could be obtained as follows

$$\gamma_{phys} = \frac{\gamma_{lbm}\, m_o}{t_o^2}, \tag{44}$$

where, $\gamma_{lbm}$ is the lattice interfacial tension, and it is measured in $(m.u./t.u.^2)$, and $\gamma_{phys}$ is the interfacial tension, and it is measured in Newtons per meter $N/m$.

### F. MF-LBM-CUDA

The simulations shown in this work used the MF-LBM C++/CUDA version of the MF-LBM solver [22]. The original MF-LBM solver [20] [44] was written on Fortran 90 and employed an MPI-OpenACC/OpenMP hybrid programming model, supporting only double-precision computations. Moreover, it supports only Linux operating systems. The solver was rewritten using the C++ language and the Nvidia C++/CUDA interface. It supports single and double-precision computations. It also supports Microsoft Windows and Linux operating systems. Single precision was added to benefit from the high computational power of Nvidia GPUs used in consumer-grade computers. Such devices are becoming more powerful and equipped with many single-precision computational units. All simulations in this work were computed with single precision using an Nvidia RTX 3090 GPU. For the porous medium simulations, the solver achieved a computational speed of 2800 – 3000 MLUPS (Million Lattice Updates Per Second).

## III. Validation

We start by validating the CGM model using two benchmark cases: the Laplace law test and the static contact angle test, where in both benchmarks, we compare the simulation output values against the input values provided to the model.

### A. Laplace law

We carried out this test to verify that the model obeys the Laplace law in three dimensions $\left(\Delta P = \frac{2\gamma}{R}\right)$, where ΔP is the difference between the pressure inside and outside the droplet, γ is the interfacial tension, and R is the droplet radius. The simulation setup comprised placing a droplet of fluid "r" inside a fully periodic domain of fluid "b". The domain size was $128\ x\ 128\ x\ 128\ l.u.$, and the initial droplet size was varied. The length unit was assumed to be $(a_o = 1\ \mu m/l.u)$. Both fluids were assumed to have equal density $\left(\rho_{phys}^r = \rho_{phys}^b = 1000\ \ kg/m^3\right)$. The lattice reference density was assumed as $(\rho_o = 1\ \ m.u/l.u^3)$. The mass unit is evaluated using Eq. (41) as $(m_o = 1 \times 10^{-15})$. The viscosity ratio between the two fluids was $\left(M = \frac{\nu^r}{\nu^b} = 5\right)$ by using the following relaxation times: $\tau^r = 1.0, \tau^b = 0.6$. This could be assumed to be corresponding to the following lattice fluid kinematic viscosities using Eq. (13):

$$\nu^r_{lbm} = \frac{1}{6}\ l.u^2/t.u,$$

$$\nu^b_{lbm} = \frac{1}{30}\ l.u^2/t.u.$$

The physical fluid kinematic viscosities are assumed as follows:

$$\nu^r_{phys} = 5 \times 10^{-6}\ m^2/s,$$

$$\nu^b_{phys} = 1 \times 10^{-6}\ m^2/s.$$

Hence, the time unit could be evaluated using Eq. (42) as $\left(t_o = \frac{1}{30} \times 10^{-6}\right)$. The physical interfacial tension between the two fluids was assumed as $(\gamma_{phys} = 0.02\ N/m)$, which was converted to the lattice interfacial tension value using Eq. (44) and set as $\left(\gamma_{lbm} = \frac{1}{45}\ m.u./t.u.^2\right)$.

After the simulations reached a steady state, the pressure inside and outside the droplet was obtained and plotted against twice the inverse of the obtained droplet radius, as shown in Figure 1. The pressure values were converted from the lattice units to the physical units using Eq. (43). It could be noted that the Laplace test is satisfied as the relationship between twice the inverse of the droplet radius and the pressure drop across the interface was linear. The magenta dotted line represents the data's linear regression, giving an interfacial tension (slope) value of $0.02009\ N/m$. The error percentage between the input interfacial tension value to the model $(\gamma_{input})$ and the simulation value of interfacial tension $\left(\gamma_{simulation} = \frac{\Delta P}{2\,R}\right)$ could be evaluated as $\left(error\,\% = \frac{|\gamma_{simulation} - \gamma_{input}|}{\gamma_{input}} \times 100 = 0.45\,\%\right)$. The very low error percentage shows the excellent agreement between the simulation model and the benchmark in capturing the interfacial tension between the two fluids.

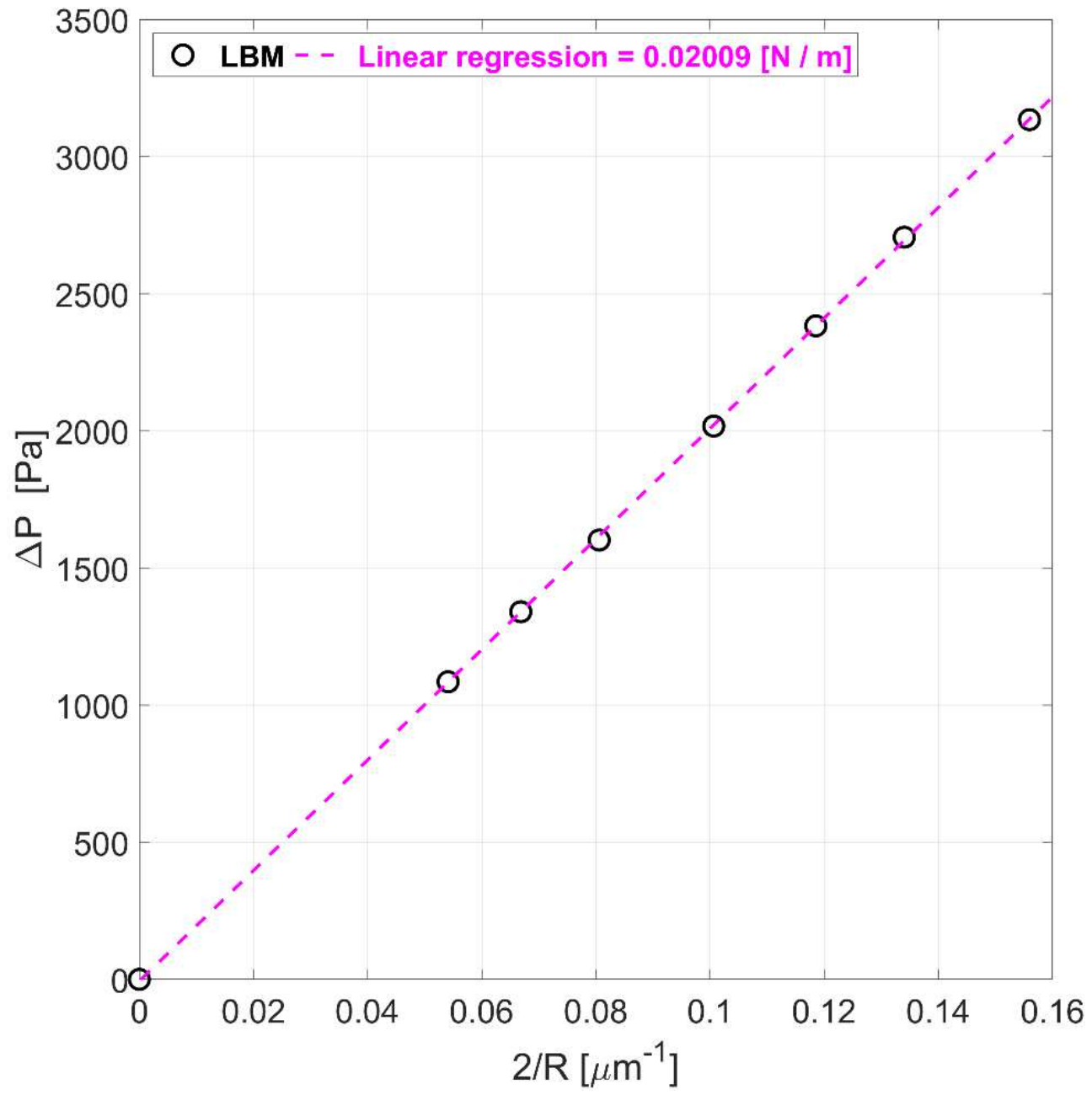

**Figure 1 Laplace law test results. The dotted line in magenta represents the linear regression between $\Delta P$ and $\frac{2}{R}$ values.**

**B. Static contact angle test**

This test was carried out to validate the input contact angle $(\theta_{input})$ to the CGM wetting model against the resulting contact angle in simulations $(\theta_{simulation})$. The contact angles extracted from the simulations using the two wetting models shown in Sec. II.B were compared against the input contact angles. The simulation setup comprised placing a droplet of fluid "r" inside a periodic fluid channel of fluid "b", where solid plates were located at the top and bottom of the channel, and the channel was periodic in the other directions. The domain size was 102 x 201 x 201 lattice points, and the droplet radius was 30 $l.u$. Other simulation parameters were set as shown in Sec. III.A.

After the simulations reached a steady state, the contact angle was obtained using a geometrical technique, where the base and height of the fluid droplet were determined using a custom *"Python Calculator"* filter in the ParaView software [45]. In that filter, a *"Plot Over Line"* source is used to identify the droplet's base or height values via a two-dimensional slice of the phase order parameter ($\phi$) field. For instance, the two locations along the base of the droplet where $\phi$ changes from a positive value to a negative value represent the two interface points between the fluids (where $\phi$ is approximately zero), which gives the droplet's base length value. The same criterion was used to determine the droplet's height. The contact angle could be determined using the droplet's base and height as follows [46]

$$\theta = \arctan\left(\frac{b/2}{r-h}\right) \tag{45}$$

$$r = \frac{4h^2 + b^2}{8h} \tag{46}$$

where, $b$ and $h$ are the droplet's base and height, respectively.

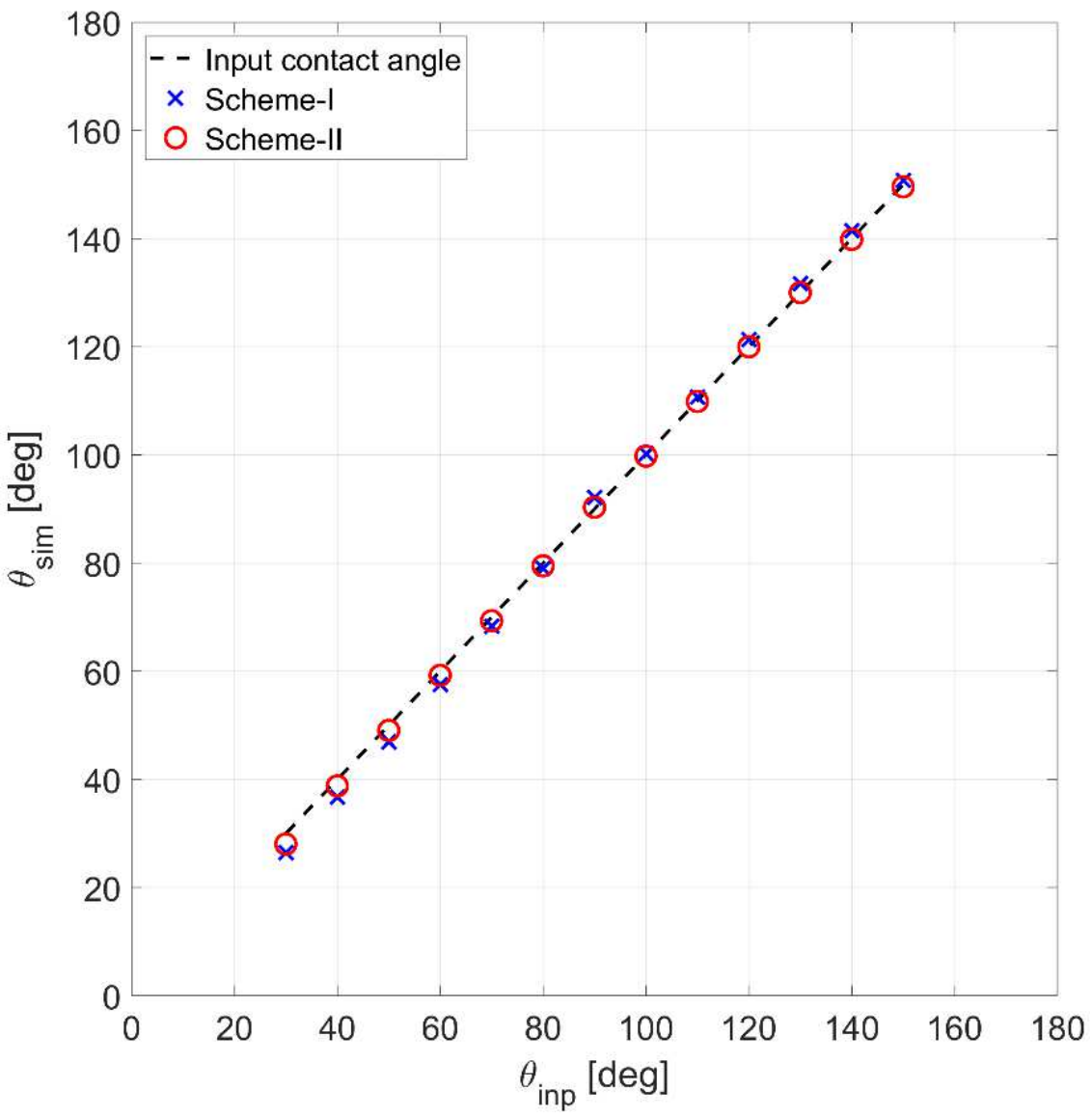

**Figure 2 Comparison between input contact angles $(\theta_{inp})$ and simulation contact angles $(\theta_{sim})$ for two wetting models: Scheme-I and Scheme-II .**

The simulated and input contact angles were compared for the two wetting models, as shown in Figure 2. It could be noted that both models had an overall excellent agreement with the input contact angle. The simulated contact angle range covers most of the practical applications. The difference between the input contact angle and the simulated contact angle $\left(\theta_{err} = \theta_{input} - \theta_{simulation}\right)$ is shown in Figure 3. Both wetting models have a higher error at lower contact angles than higher ones. Moreover, Scheme-II showed better accuracy in this test, producing less error in most cases. It also produced mostly positive error values. However, the largest errors for Scheme-I were not very high either, and they were within 3.6° and -2.2°. The errors for Scheme-I were positive for lower contact angles ($\theta < 90°$) and negative for higher contact angles ($\theta > 90°$). Notably, both schemes produced odd errors at ($\theta = 90°$) that did not follow the error trend of the scheme. This could be attributed to the mathematical expressions of the models where the contact angle of 90° will result in the mathematical expression ($\cos(90°) = 0$), which results in special conditions in the equations and even leads to entirely neutralizing Scheme-I. Moreover, it should be noted that the implementation of Scheme-I in the MF-LBM solver has a limitation of setting the contact angle as ($\theta < 90°$), while Scheme-II does not have this limitation. Since the contact angle is assumed to be always measured through fluid "b", this implies that fluid "r" will always be the non-wetting fluid while fluid "b" will always be the wetting fluid in the MF-LBM solver. However, setting a contact angle higher than 90° using Scheme-I was achieved by switching the positions of the fluids and changing the assumption that the contact angle was measured through the fluid "r" instead of "b". For instance, setting a contact angle ($\theta = 30°$) was achieved by setting the fluid "r" as the fluid inside the droplet while setting fluid "b" as the surrounding fluid and measuring the contact angle through the fluid "b". On the contrary, setting a contact angle ($\theta = 150°$) was achieved by setting the fluid "b" as the fluid inside the droplet while setting fluid "r" as the surrounding fluid and measuring the contact angle through the fluid "r". This is shown in Figure 4. Those switches were not necessary for Scheme-II as it could freely achieve contact angles higher and lower 90°.

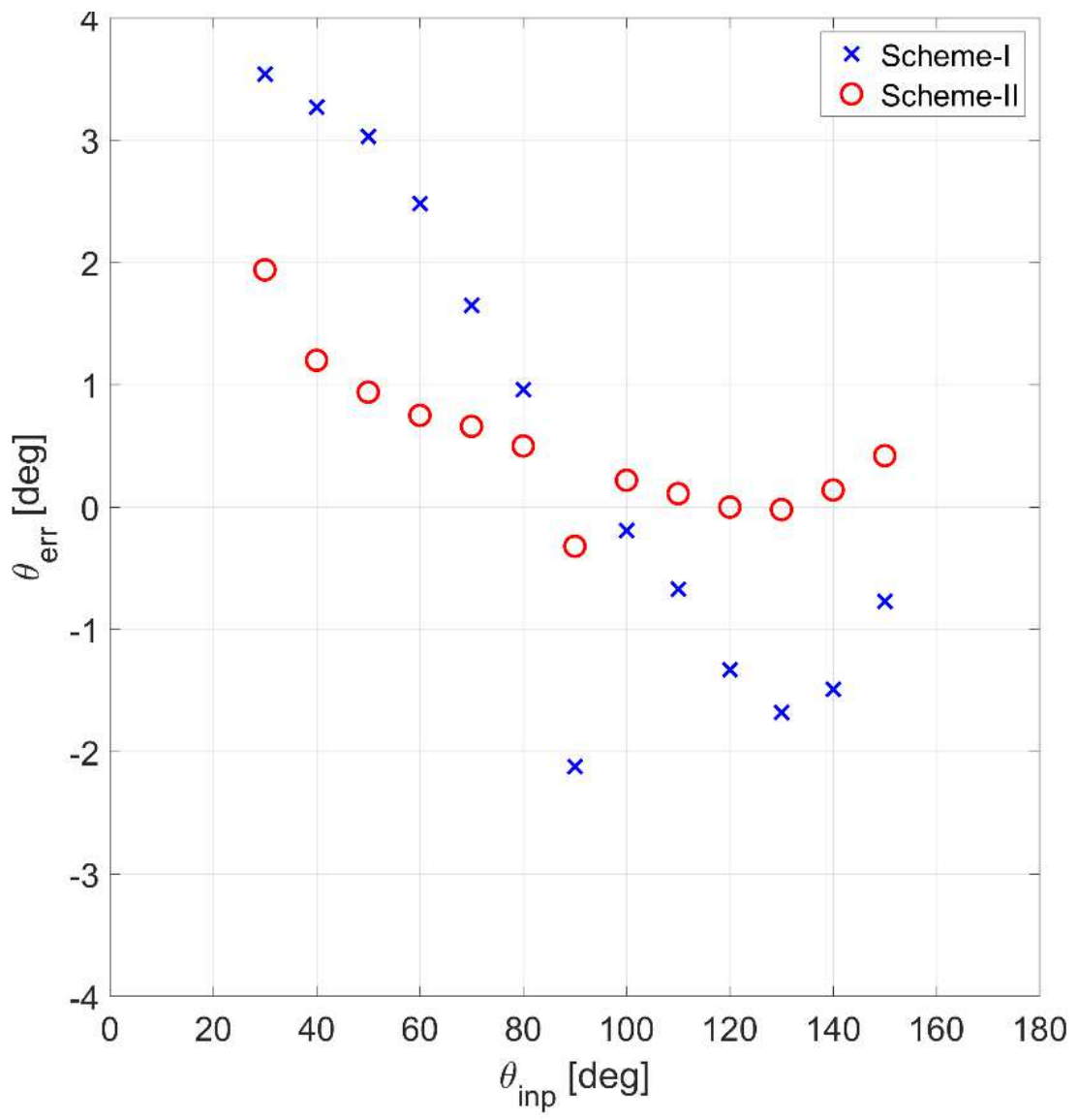

**Figure 3 Error between input contact angle and simulated contact angle in degrees at each input contact angle value.**

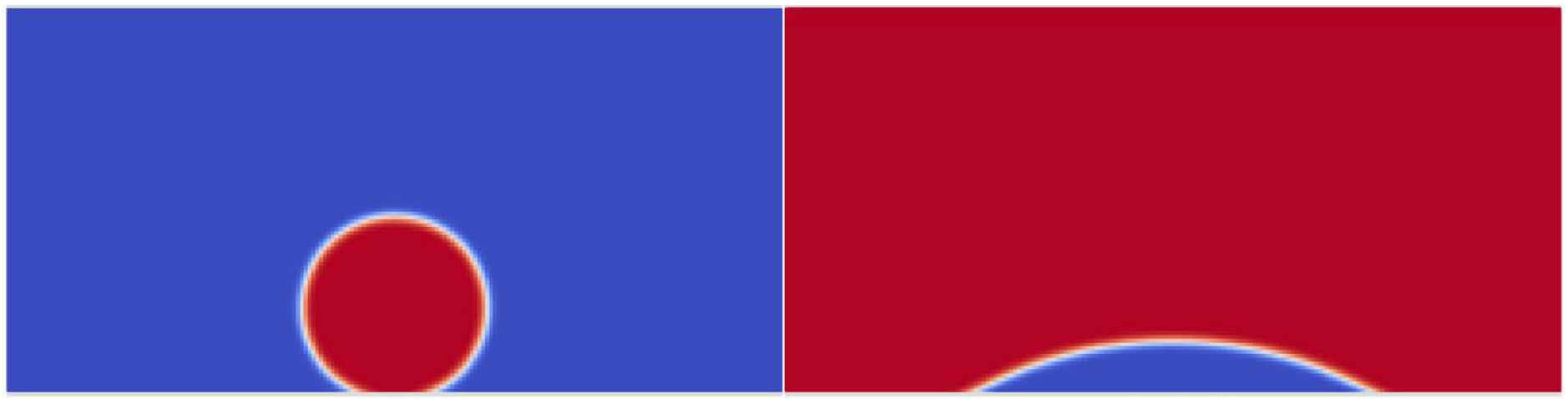

**Figure 4 Comparison between the layout of the fluids for contact angles lower and higher than 90° using Scheme-I of wetting models. The snapshots represent the phase order parameter field ($\phi$) ranging between red for the value (1) and blue for the value (-1) and representing positions of fluid "r" and "b", respectively.**

## IV. Numerical simulation results

### A. Binary fluid flow in a horizontal capillary tube

We simulated the displacement processes of two fluids in a cylindrical tube where the Washburn equation of multiphase flow in a single capillary tube applies [33]. The Washburn equation – after ignoring the slip at walls – is given by

$$z^2 \frac{\mu_w - \mu_{nw}}{2} + \mu_{nw}\, L\, z = \frac{r^2}{8}\left(\Delta P + \frac{2\,\gamma}{r}\cos(\theta)\right) t, \tag{47}$$

where, $z$ is the interface tip position in the tube, $r$ is the tube radius, and $t$ is the time the fluid interface takes to reach the position $z$. The above equation becomes:

$$z(t) = -\frac{L\,\mu_{nw}}{\mu_w - \mu_{nw}} + \sqrt{\left(\frac{L\,\mu_{nw}}{\mu_w - \mu_{nw}}\right)^2 + \frac{1}{2\,(\mu_w - \mu_{nw})}\left(\frac{\Delta P\, r^2}{2} + r\,\gamma\,\cos(\theta)\right) t}, \tag{48}$$

which could be rewritten in a non-dimensional form as follows:

$$z^*(t) = -\left(\frac{\mu_{nw}}{\mu_w - \mu_{nw}}\right) + \sqrt{\left(\frac{\mu_{nw}}{\mu_w - \mu_{nw}}\right)^2 + \frac{1}{2}\left(\frac{\Delta P\, r^2}{2\,L\,\gamma} + \frac{r}{L} cos(\theta)\right) t^*}, \tag{49}$$

where, the non-dimensional length ($z^*$) and non-dimensional time ($t^*$) are defined as

$$z^* = \frac{z}{L}, \tag{50}$$

$$t^* = \frac{\gamma}{(\mu_w - \mu_{nw})\, L}\, t. \tag{51}$$

Two displacement processes were simulated where Eq. (49) could be used as an analytical solution to verify the numerical simulation results. The first case represented a spontaneous imbibition displacement process, where the fluids moved under the wettability effect ($\theta < 90°$) and there was no pressure difference between the inlet and the outlet of the tube $\Delta P = 0$. In the second case, it was assumed that both fluids were neutrally wetting ($\theta = 90°$) and the fluids moved under the effect of a pressure difference between the inlet and the outlet of the tube ($\Delta P > 0$).

The simulation setup consisted of a tube with a radius of 5 $l.u.$ and the domain dimensions were $12 \times 12 \times 101\ l.u.$ This coarse mesh is comparable to that in the small pores in porous media that will be simulated later. This was vital to assess the model's capabilities in simulating the displacement processes in a small pore with a coarse mesh. The length unit was assumed as $(a_o = 1 \times 10^{-6}\ m/l.u)$. That represented a capillary tube with a radius of $5\ \times 10^{-6}\ m$ and length $100\ \times 10^{-6}\ m$ as shown in Figure 5. The fluid "b" (colored blue in Figure 5) was placed at the inlet of the tube, and it represented the wetting fluid, while the fluid "r" (color red in Figure 5) was placed at the rest of the tube, and it represented the non-wetting fluid. Both fluids were assumed to have equal density $\left(\rho^r_{phys} = \rho^b_{phys} = 1000\ kg/m^3\right)$. The lattice reference density was assumed as $(\rho_o = 1\ m.u/l.u^3)$. The mass unit is evaluated using Eq. (41) as $(m_o = 1 \times 10^{-15})$. The viscosity ratio between the two fluids was $\left(M = \frac{\nu^r}{\nu^b} = \frac{1}{5}\right)$ by using the following relaxation times: $\tau^b = 1.0, \tau^r = 0.6$. This could be assumed to be corresponding to the following lattice fluid kinematic viscosities using Eq. (13):

$$\nu^b_{lbm} = \frac{1}{6}\ l.u^2/t.u,$$

$$\nu^r_{lbm} = \frac{1}{30}\ l.u^2/t.u.$$

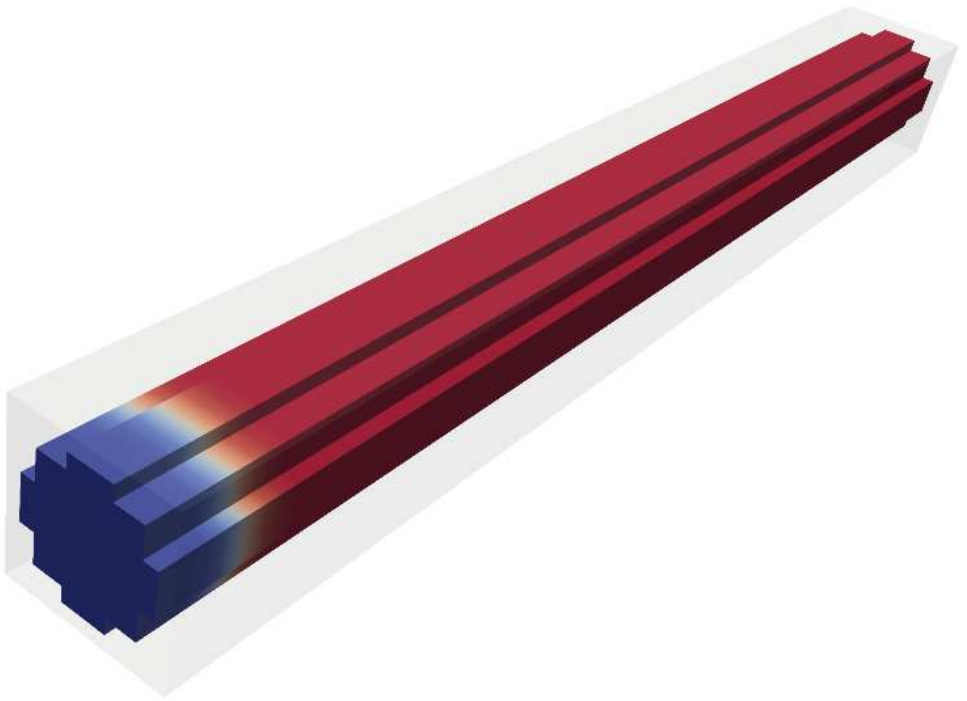

**Figure 5 Simulation setup for two phase flow in a horizontal capillary tube.**

The physical fluid kinematic viscosities are assumed as follows:

$$\nu^b_{phys} = 5 \times 10^{-6}\ m^2/s,$$

$$\nu^r_{phys} = 1 \times 10^{-6}\ m^2/s.$$

Hence, the time unit could be evaluated using Eq. (42) as $\left(t_o = \frac{1}{30} \times 10^{-6}\right)$. The physical interfacial tension between the two fluids was assumed as $\left(\gamma_{phys} = 0.02\ \ N/m\right)$, which was converted to the lattice interfacial tension value using Eq. (44) and set as $\left(\gamma_{lbm} = \frac{1}{45}\ m.u./t.u.^2\right)$. The initial conditions for all simulation cases were determined by running an initial simulation where the fluid "b" (neutrally wetting) was placed at only the first layer of the tube, while the rest of the tube was filled with the fluid "r", and the simulation was run until a steady-state interface between the two fluids was achieved as shown in Figure 5. After running the initial simulation, the initial position of the interface was approximately $\left(z_{init}^{*}\sim\frac{5.4}{100}\right)$ which corresponds to an initial time $(t_{init}^{*})$ using Eq. (49), depending on the simulated contact angle or pressure drop. The initial time was considered in the comparison between the simulation results and Eq. (49). The boundary conditions at both the inlet and the outlet sides of the tube were set as pressure boundary conditions (as shown in Sec. II.C.1) using equal pressures at the inlet and the outlet. The wetting fluid was injected into the tube's inlet using the technique mentioned in Sec. II.D.1. The Washburn equation assumes a constant contact angle at the flow meniscus throughout the fluid flow. Therefore, the dynamic contact angle was not considered when comparing the simulation results and the analytical solution of the Washburn equation.

### 1. Case-I: Spontaneous imbibition displacement

In this case, the simulation results using the two wetting models shown in Sec. II.B are compared against the Washburn equation (Eq. (49)). Different contact angles were simulated and compared, as shown in Figure 6 and Figure 7. All the simulated contact angles were below 90° to mimic an imbibition displacement process where the invading wetting fluid at the inlet boundary displaces the nonwetting fluid inside the tube. It could be noted from the figures that both wetting model schemes provided a very good agreement with the analytical solution of the Washburn equation at different contact angles. Both wetting model schemes captured the system's underlying physics, where the interface between the two fluids moved rapidly in the fluid displacement, and then it slowed down as it proceeded through the tube. Also, the fluid interface moved faster at lower contact angles due to higher wettability. However, it could be noted from Figure 8 that Scheme-I was mostly more accurate than Scheme-II as it produced less relative error at most contact angles $\left(error\ \% = \frac{\left|\int z_{lbm}(t)dt - \int z_{analytical}(t)dt\right|}{\int z_{analytical}(t)dt} \times 100\right)$. The error was evaluated as the relative difference between the areas under the curves of both the simulation results and the analytical solution. At lower contact angles, both schemes produced relatively larger errors. This could be attributed to the fast dynamics of the fluid's movement as the more viscous fluid "b" was strongly wetting, and the fluid interface moved rapidly due to the wettability. Under such conditions, the negative impact of the initial interface position on the simulation results was also increased. Moreover, it could be noted that the error produced by Scheme-II kept decreasing as the contact angle was increased. As for Scheme-I, for most contact angles, the error decreased as the contact angle was increased, except for $\theta = (70°, 80°)$. Snapshots of the simulation results at $t^* \approx 50$ at different contact angles are shown in Figure 9. The snapshots show a slice of the phase field order parameter $(\phi)$ along the middle of the tube, where the fluid "b" (colored in blue) entered the tube from the left side and displaced the fluid "r" (colored in red). It could be concluded from this test that Scheme-I is more accurate in simulating dynamic movement of fluids interface under wettability effects.

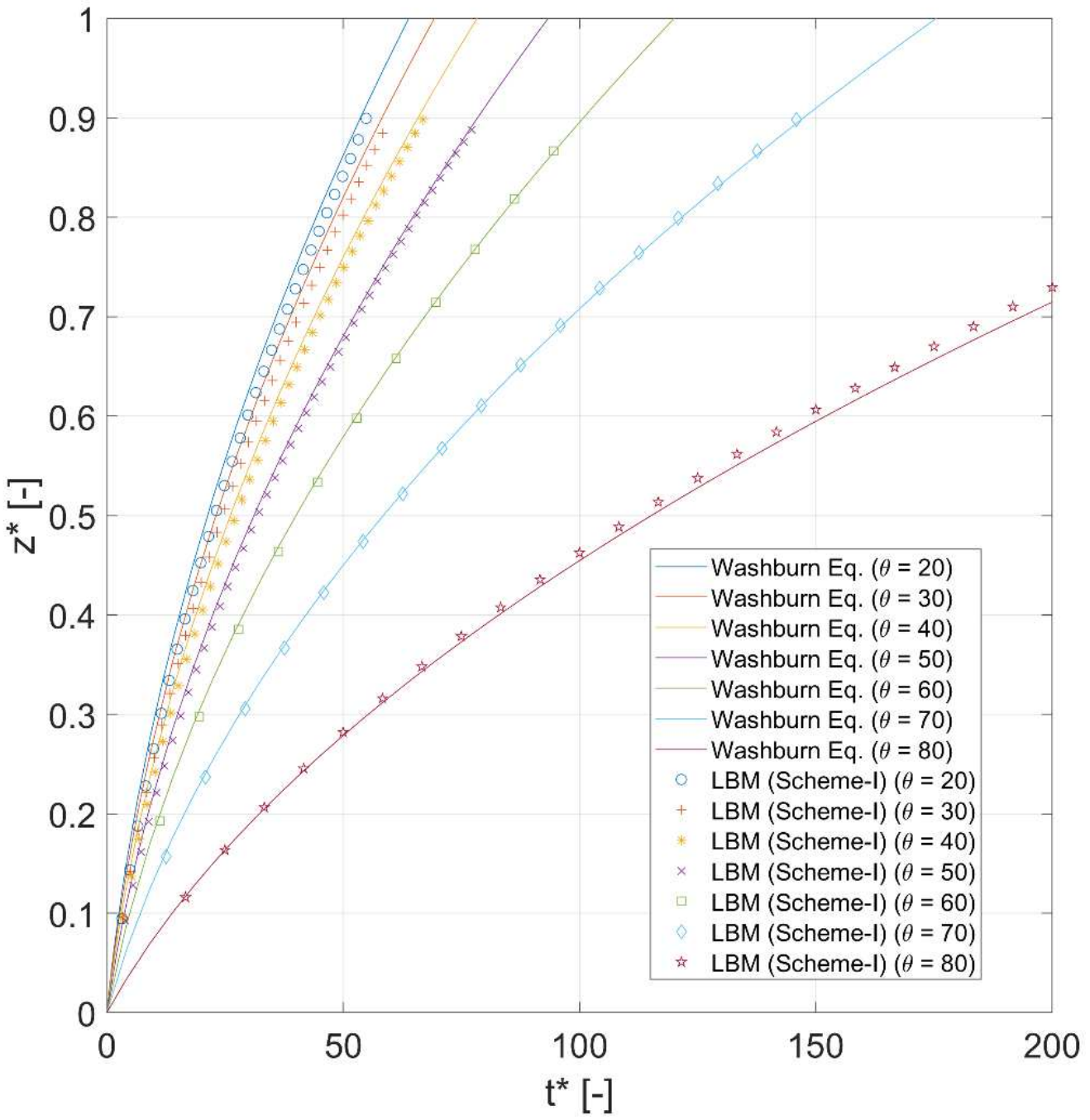

**Figure 6 Comparison between the Washburn equation and simulation results using Scheme-I of wetting models.**

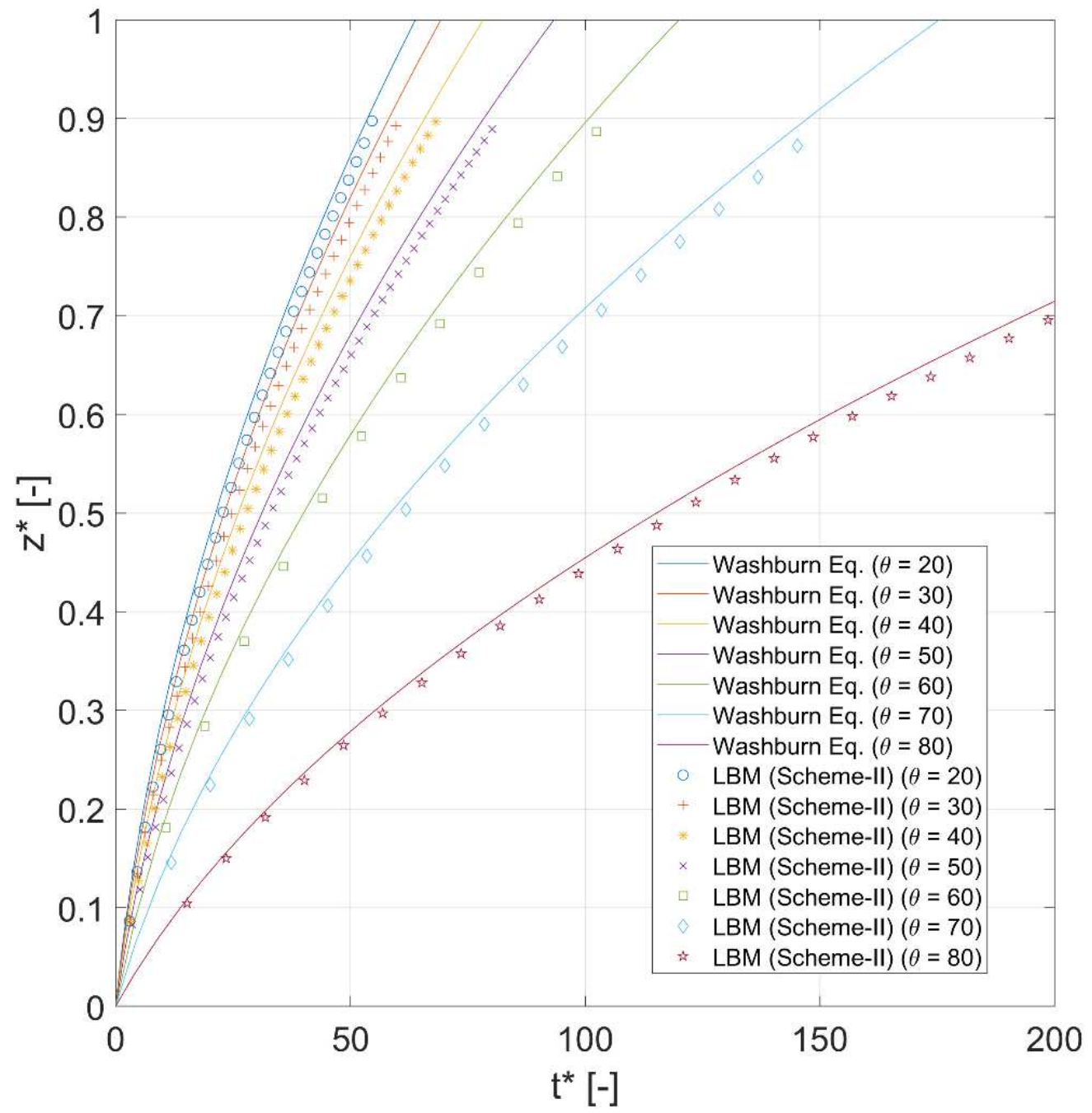

**Figure 7 Comparison between the Washburn equation and simulation results using Scheme-II of wetting models.**

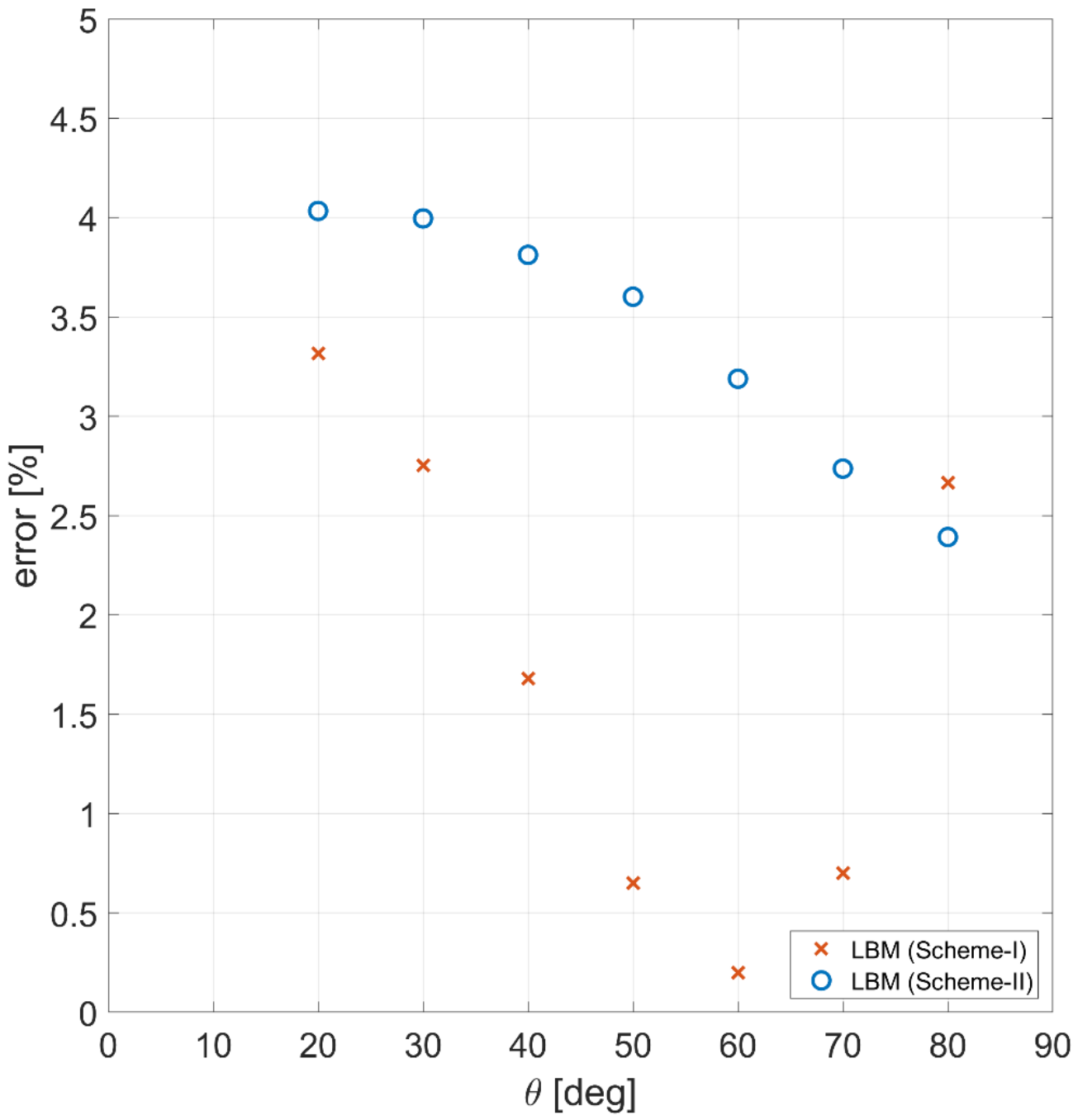

**Figure 8 Relative error percentage between simulation results using wetting model schemes and the analytical solution of the Washburn equation.**

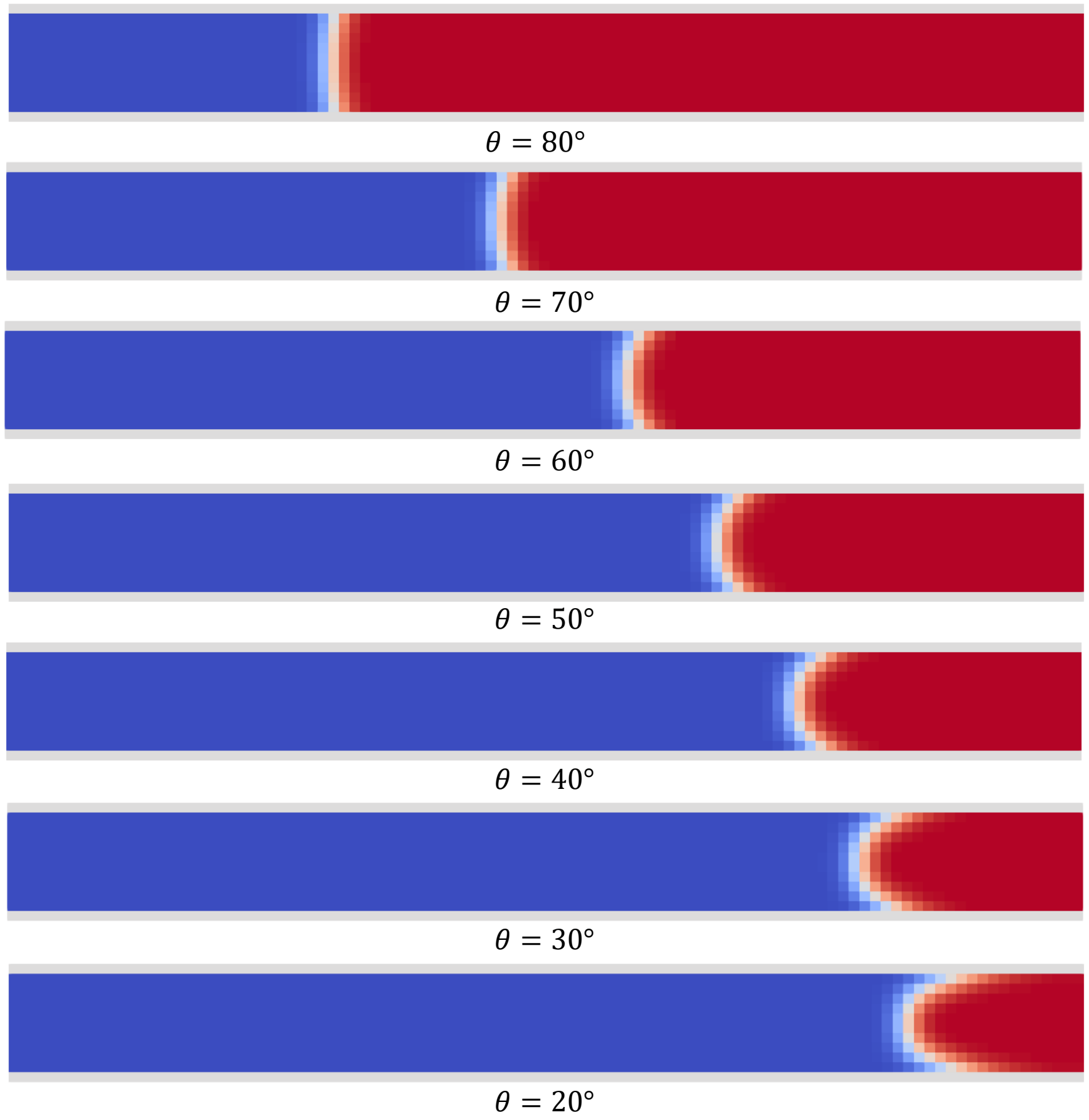

**Figure 9 Snapshots of fluid interface advancement in the tube. The snapshots were taken at $t^* \approx 50$ at different contact angles using Scheme-II for a slice passing through the center of the tube.**

**2. Case-II: Pressure gradient displacement**

In this case, the simulation results using two pressure gradient implementations are compared against the Washburn equation (Eq. (49)). Several pressure differences between the tube's inlet and outlet were simulated and compared, as shown in Figure 10 and Figure 11. For both implementation types, the numerical simulation results deviated slightly from the analytical solution of the Washburn equation, especially at higher pressures. However, both implementations captured the underlying physics properly, where the fluids interface moved more rapidly at higher pressures, and it quickly moved at first, then slowed down as it progressed through the tube. It could be noted from Figure 12 that Type-I implementation provided slightly more accurate results for most pressure drops. At higher pressure differences, both types produced relatively larger errors. This could be attributed to the fast dynamics of the fluid's movement at higher pressures. Moreover, it could be observed that at low-pressure differences, Type-I proved much more accurate results than Type-II. Still, the accuracy difference was not very high for most pressure differences. It could be concluded from this test that both pressure gradient implementation types have nearly the same accuracy, with Type-I slightly more accurate.

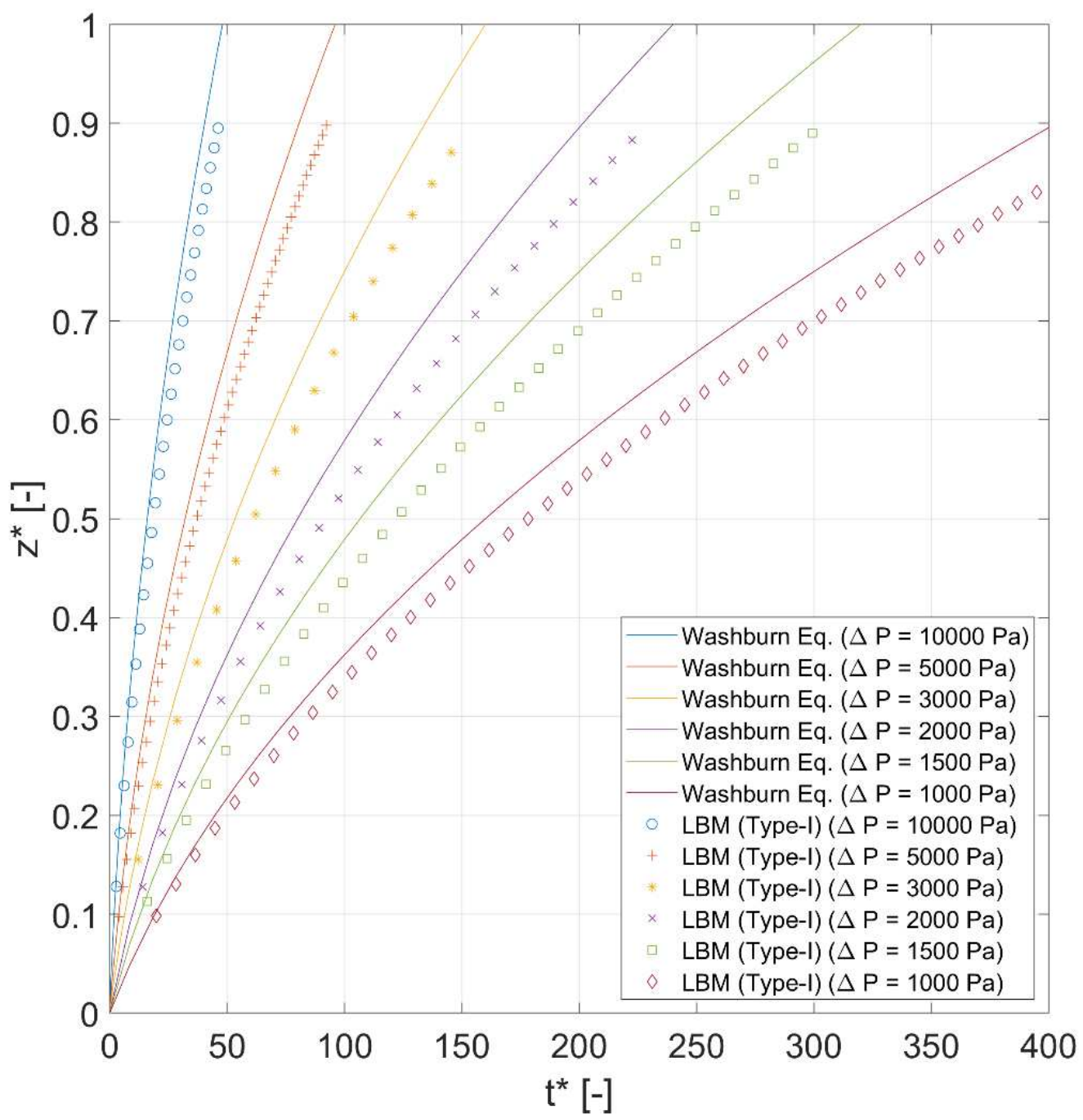

**Figure 10 Comparison between the Washburn equation and simulation results using Type-I of pressure gradient implementation.**

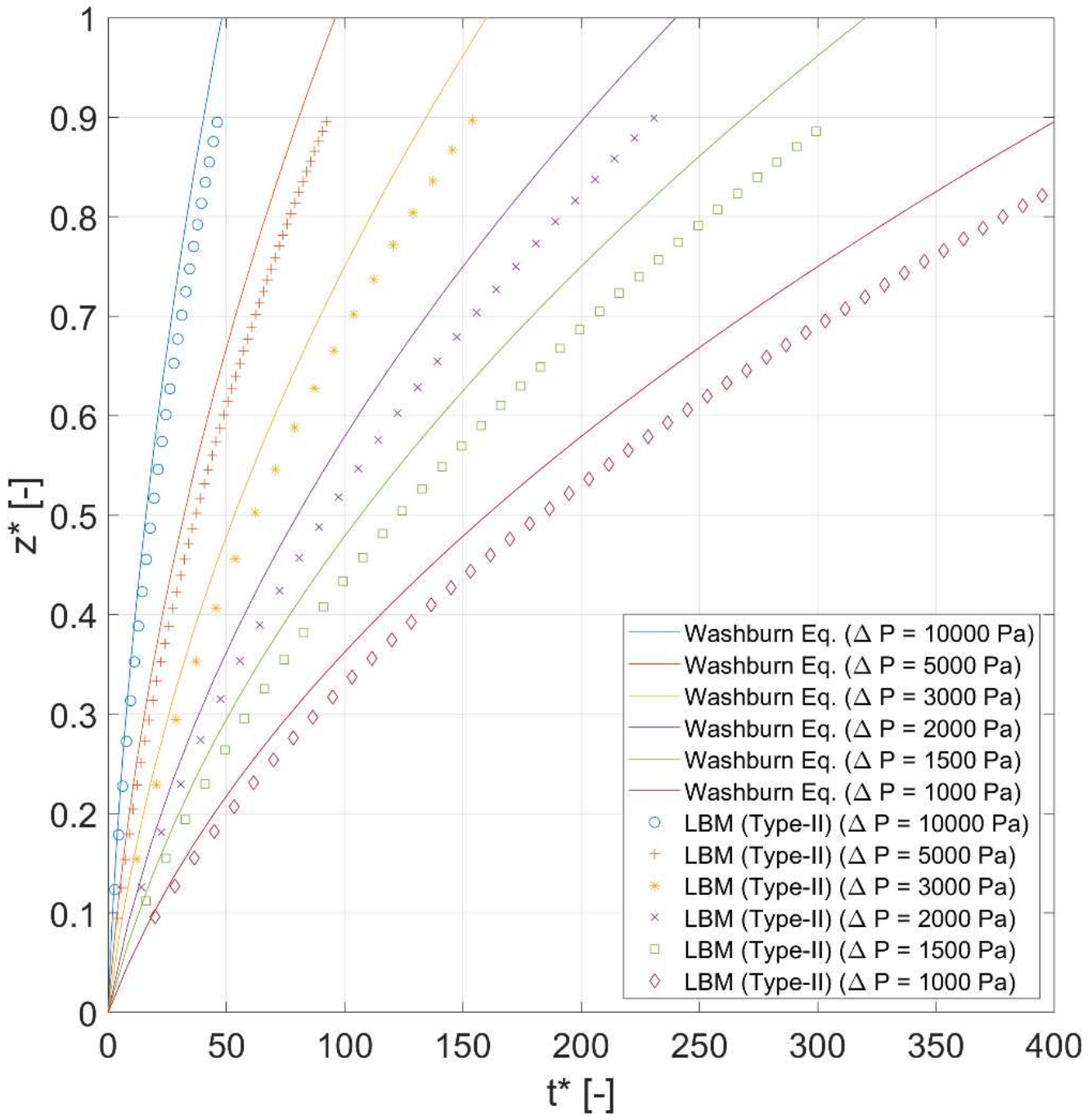

**Figure 11 Comparison between the Washburn equation and simulation results using Type-II of pressure gradient implementation.**

It could also be noted that the errors between the simulation results and the analytical solution of the Washburn equation for fluid movement under pressure gradient are larger than in simulations for fluid movement for different contact angles without any pressure gradient. One of the contributing factors to this higher error is the error in the effective contact angle, as the two fluids are not exactly neutrally wet. For instance, at $\Delta P = 10000\ Pa$ implemented by either Type-I or Type-II, the error percentage was $(\approx 9\ \%)$. However, if the input contact angle to the Washburn equation was slightly altered from $(\theta = 90°)$ to $(\theta = 92°)$ – similar to the errors observed in the contact angle test in Sec. III.B at $(\theta = 90°)$ without a wetting model – the error percentage was reduced to $(\approx 7\ \%)$. Moreover, setting the contact angle in the Washburn equation to $(\theta = 100°)$ made the error percentage drop to negligible values $(\approx 0.25\ \%)$. It illustrates that the effective contact angle described by the Washburn equation might differ slightly from the imposed one.

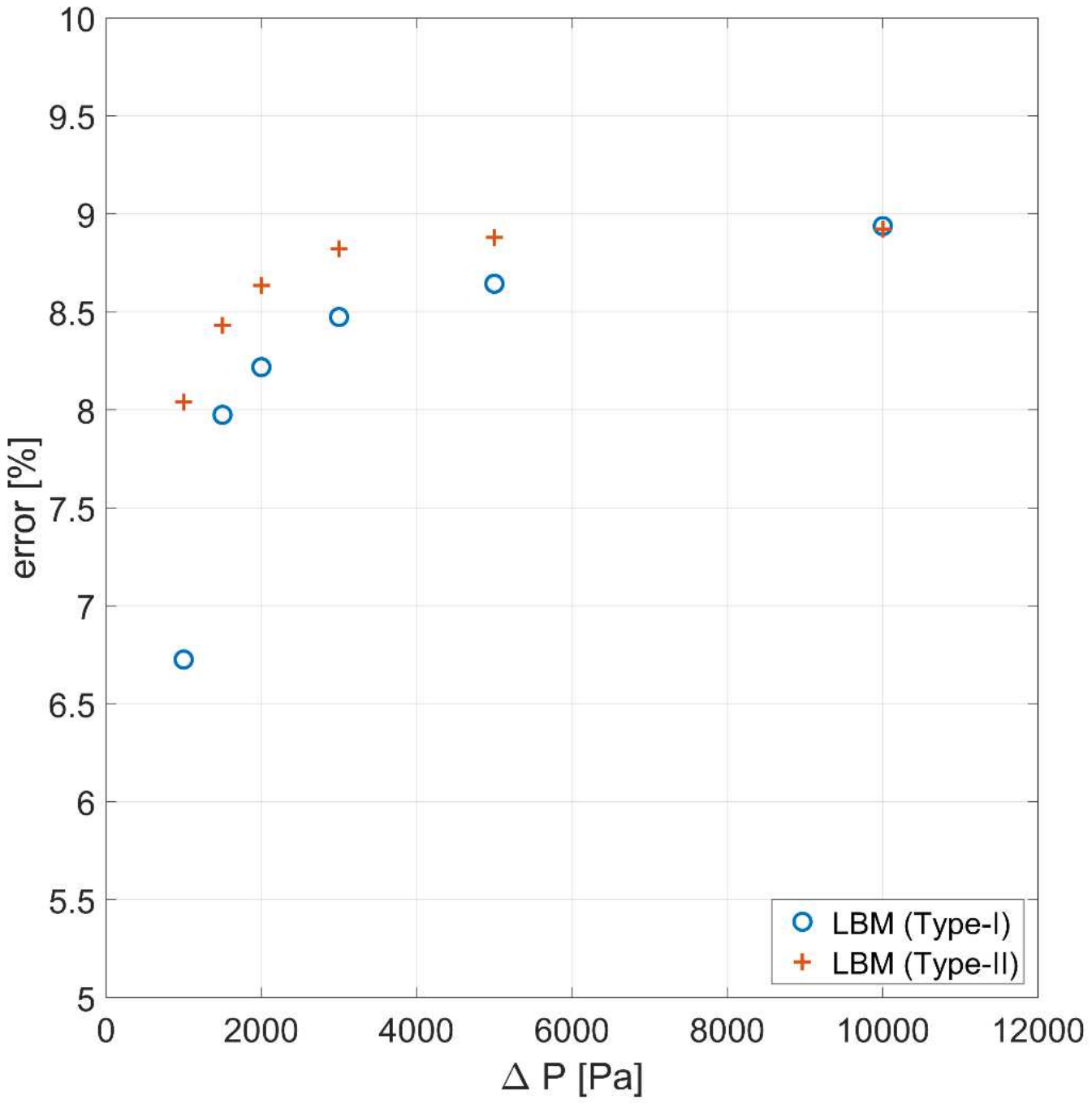

**Figure 12 Relative error percentage between simulation results using wetting model schemes and the analytical solution of the Washburn.**

**B. Realistic Porous Medium**

In this section, we simulated two displacement processes encountered in petroleum reservoirs: primary drainage and imbibition. These processes are normally characterized by laboratory experiments known as special core analysis (SCAL) experiments [47]. However, those experiments are lengthy and expensive. Appropriate LBM simulations have the potential to mimic the SCAL experiments and provide useful results promptly. We conducted the simulations using the pressure gradient implementation in Sec. II.C and compared the difference between the simulation results for each implementation.

The simulation setup comprised a sample from a porous medium, a Bentheimer sandstone rock. The geometry was extracted from one of the Digital Rocks portal projects named "*A large scale X-Ray micro-tomography dataset of steady-state multi-phase flow*" [48]. The original size of the dataset was 1950 x 1950 x 10800 voxels with a voxel resolution of 6 µm. Thus, the original dataset size was 11700 µm x 11700 µm x 64800 µm, representing a core sample from a

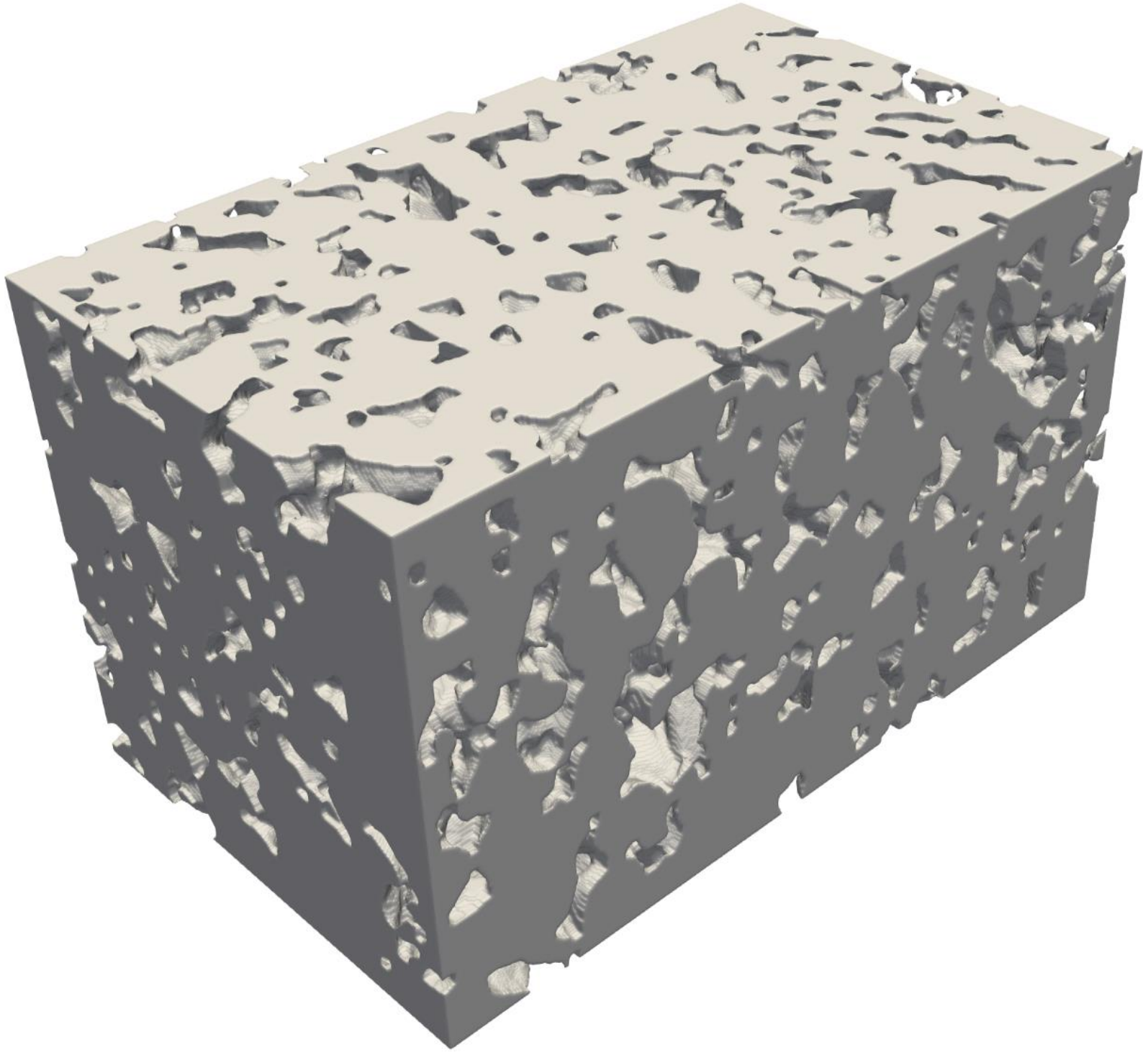

**Figure 13 An isometric view of the Bentheimer sandstone sample used in the simulations.**

Bentheimer sandstone. A subset with the size of 300 x 300 x 500 voxel (1800 µm x 1800 µm x 3000 µm) was extracted and used in the present simulations. The extracted sub-sample was pre-processed before being used in the simulations, as shown in previous work [3]. The isolated pores that do not contribute to the fluid flow were removed from the rock sample using the MATLAB software image processing functions such as "*bwareaopen*" [49]. Also, very narrow pores were removed from the geometry MATLAB function "*imclose*" [50]. The processed geometry is shown in Figure 13. The sample porosity was 21.3%.
Solid walls were added to all faces of the sample except for the inlet and outlet faces. The setup mimicked the conditions in laboratory experiments where the core sample is fixed in a solid core holder and surrounded by confined pressure to force the fluid movement laterally from the inlet to the outlet and avoid escaping through the core sides. Also, ten layers of pure fluid lattice points were added at the sample's inlet and outlet to facilitate injecting the inlet fluid and receiving the outlet fluids.

We assumed the simulated fluids were oil and water in a special core analysis laboratory (SCAL) experiment [47], where a primary drainage displacement process followed by an imbibition displacement process was simulated. Water was assumed to be the wetting fluid, while oil was the non-wetting fluid. The porous sample was considered to be initially filled with water. Then, oil was injected at the inlet to displace water in a primary drainage displacement. The inlet pressure was increased gradually and monitored against the water saturation in the sample. Once a large increase in the inlet pressure did not result in a large decrease in the water saturation, it was assumed that the irreducible water saturation was achieved, and the primary drainage simulation was stopped. Next, the inlet pressure was gradually decreased to allow water spontaneous imbibition into the sample. After that, water was injected into the inlet instead of the oil to simulate the forced imbibition displacement process. Once a large increase in the inlet pressure did not result in a large increase in the water saturation, it was assumed that the critical oil saturation was achieved, and the imbibition displacement simulation was stopped.

Fluid flow was driven via a pressure gradient, and the pressure gradient implementations shown in Sec. II.C were compared against each other. The wetting model Scheme-I was used in all simulations since it was shown to be superior in dynamic fluid simulations shown in Sec. IV.A.1. This simulation setup was used to mimic a widely conducted experiment in SCAL laboratories to generate the capillary pressure-water saturation $(P_c - S_w)$ curve of petroleum reservoir rocks. Since capillary pressure could be defined as the difference between the pressure in the non-wetting fluid and the wetting fluid $(P_c = P_{nw} - P_w)$, it could be assumed that the capillary pressure is the pressure difference applied as a pressure gradient in the domain. As shown in Sec. II.E, the results were converted from lattice units to physical units. The length unit was assumed to be equal to the micro-computed tomography (µCT) scan image resolution of the porous rock $(a_o = 6 \times 10^{-6}\ m/l.u)$. The fluid "r" was assumed to be oil while the fluid "b" was assumed to be water. Both fluids were assumed to have equal density $\left(\rho^r_{phys} = \rho^b_{phys} = 1000\ kg/m^3\right)$. The lattice reference density was assumed as $(\rho_o = 1\ m.u/l.u^3)$. The mass unit is evaluated using Eq. (41) as $(m_o = 2.16 \times 10^{-13})$. The physical fluid kinematic viscosities are assumed as $\left(\nu^r_{phys} = 5 \times 10^{-6}\ m^2/s\, , \nu^b_{phys} = 1 \times 10^{-6}\ m^2/s\right)$. The relaxation times were assumed as $(\tau^r = 1.0, \tau^b = 0.6)$, which gave lattice fluid kinematic viscosities as $\left(\nu^r_{lbm} = \frac{1}{6}\ l.u^2/t.u\, , \nu^b_{lbm} = \frac{1}{30}\ l.u^2/t.u\right)$. Hence, the time unit could be evaluated using Eq. (42) as $(t_o = 1.2 \times 10^{-6})$. The physical interfacial tension between the two fluids was assumed as $\left(\gamma_{phys} = 0.02\ N/m\right)$, which was converted to the lattice interfacial tension value using Eq. (44) and set as $\left(\gamma_{lbm} = \frac{2}{15}\ m.u./t.u.^2\right)$. The contact angle was set as $(\theta = 60°)$.

The simulation results are shown in Figure 14. Water saturation $(S_w)$ was obtained using the phase field order parameter defined in Eq. (7) by summing the number of fluid lattice points where $(\phi(x) < 0)$ over the total number of fluid lattice points in the domain of the porous medium. It could be noted that they qualitatively matched the typical relationship observed in laboratory experiments [47]. The primary drainage curve started with 100% wetting fluid (water) saturation. Minimum pressure was required for the nonwetting fluid (oil) to displace the wetting fluid (water) and enter the largest pores, known as the threshold pressure or displacement pressure, which was in the $500 - 1000\ Pa$ range. Then, under increasing pressure and depending on the pore-size distribution, other pores were filled with oil. As the pressure was further increased, it was noticed that changes in the water saturation were very small. Hence, the primary drainage simulation was stopped at $P_c = 5000\ Pa$.

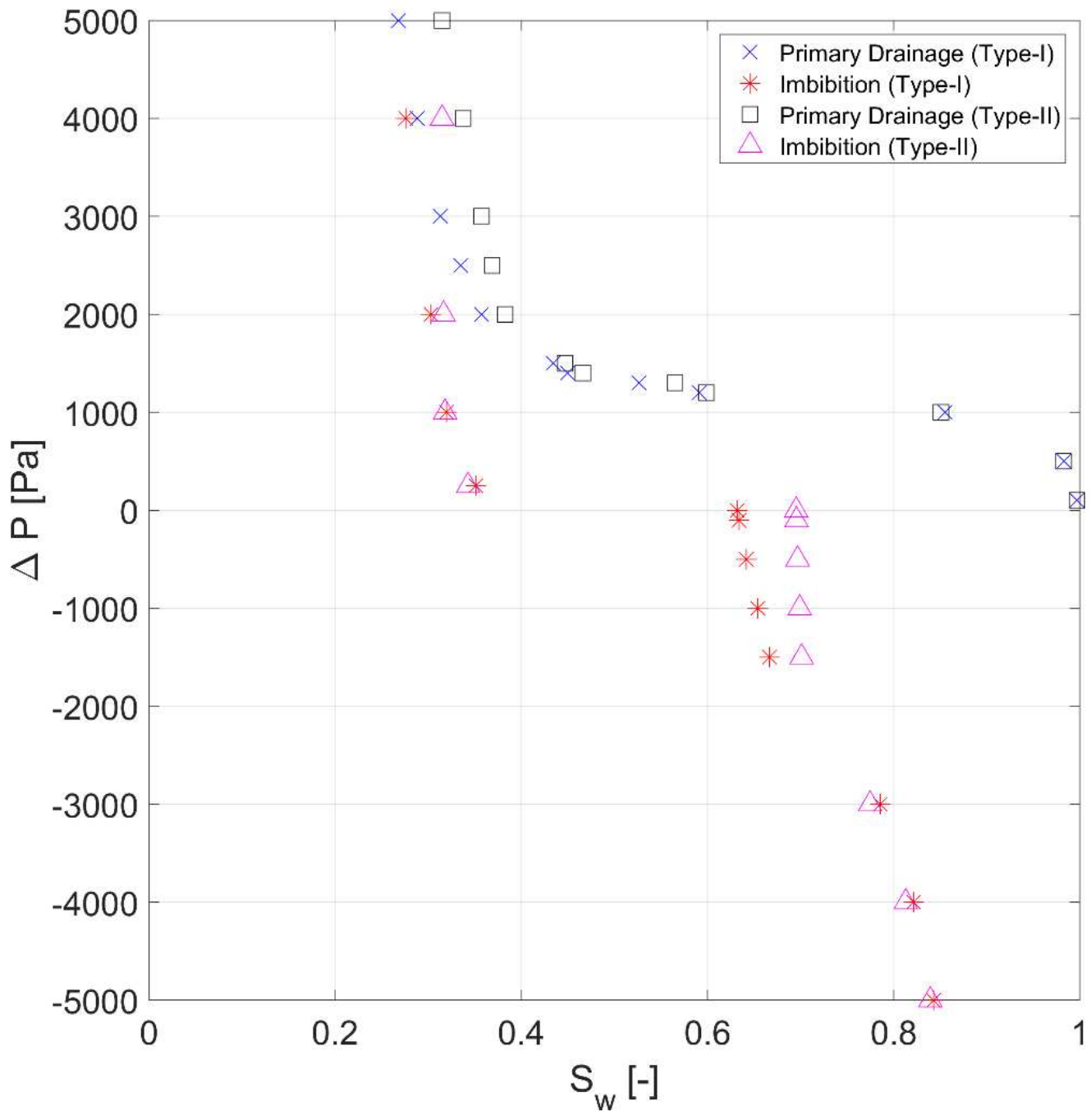

**Figure 14 Comparison between the primary drainage and imbibition simulation results for two types of pressure gradient implementations.**

The spontaneous imbibition displacement process was simulated by gradually reducing the pressure drop while keeping the oil at the inlet side of the sample, allowing the water at the outlet to return to the porous rock spontaneously. It was observed in the positive pressure difference part of the imbibition curve that water saturation was slowly increased for most pressure drops. Still, it was dramatically increased at zero pressure difference. After that, the forced imbibition displacement process was started by injecting water into the inlet of the sample. It was observed that water saturation increased slowly with the increase of the pressure difference. The pressure difference was considered negative capillary pressure since water (the wetting fluid) was injected instead of oil (the non-wetting fluid). The pressure difference was increased to -5000 Pa when it was noted that further increases in the pressure difference resulted in only small increases in the water saturation.

The pressure gradient driving the displacement processes was simulated using the two types of pressure gradient implementations shown in Sec. II.C. It could be noted that during the early stages of the primary drainage process, the results from both implementation types were nearly identical. However, at the later stage of the primary drainage process, Type-II simulations showed a lower water saturation at the same pressure difference. Lower water saturation means fewer pores were invaded using this Type-II pressure gradient implementation. The differences can be observed in the highlighted red circles in Figure 15, which shows snapshots after the end of the primary drainage simulations. The red circles at snapshots (a) & (c), and (b) & (d) are compared with each other to highlight that oil (colored in orange) invaded more pores in Type-I simulations than Type-II. This can be attributed to the differences between the two methods. While Type-II imposes a homogeneous pressure gradient along the sample, Type-I only imposes a pressure difference between the inlet/outlet. Thus, in Type-I, the direction of the pressure gradient varies locally making the invading fluid explore more pores than using Type-II.

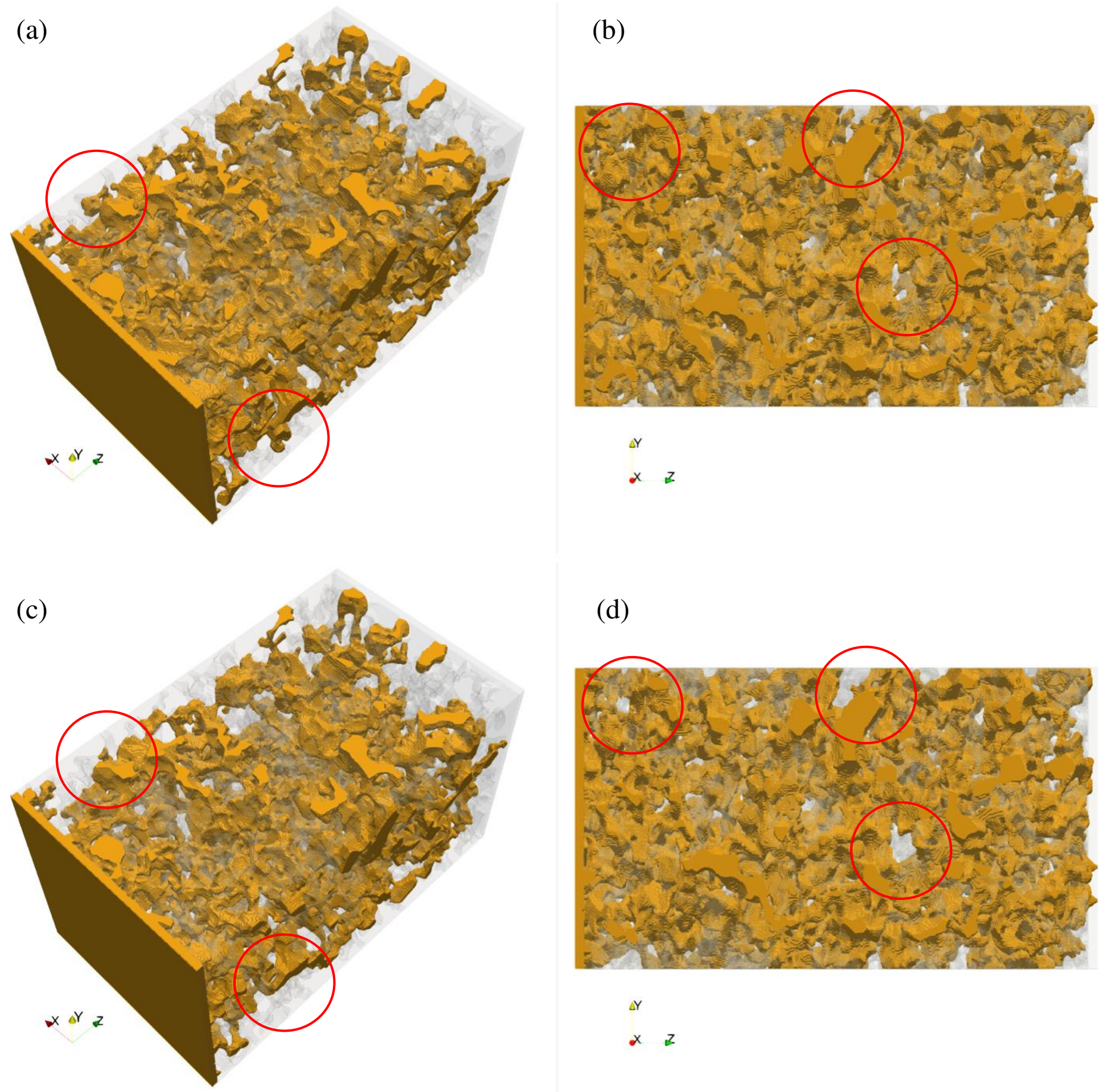

**Figure 15 Comparison between snapshots from the end of primary drainage simulations. Snapshots (a) and (b) are for Type-I implementation simulations, while snapshots (c) and (d) are for Type-II. Snapshots (a) and (c) show an isometric view of the geometry, while snapshots (b) and (d) show a front view of the geometry. Orange color represents oil, transparent grey represents solid, and water is hidden.**

The irreducible water saturation value $\left(S_{w,irr}\right)$ was different between the two types as follows: Type-I: $S_{w,irr} = 0.268$, Type-II:$S_{w,irr} = 0.315$. The values are considered within the expected range in laboratory experiments [47]. That resulted in a different starting point for the imbibition curves, which was reflected in the spontaneous imbibition water saturations as follows: Type-I:

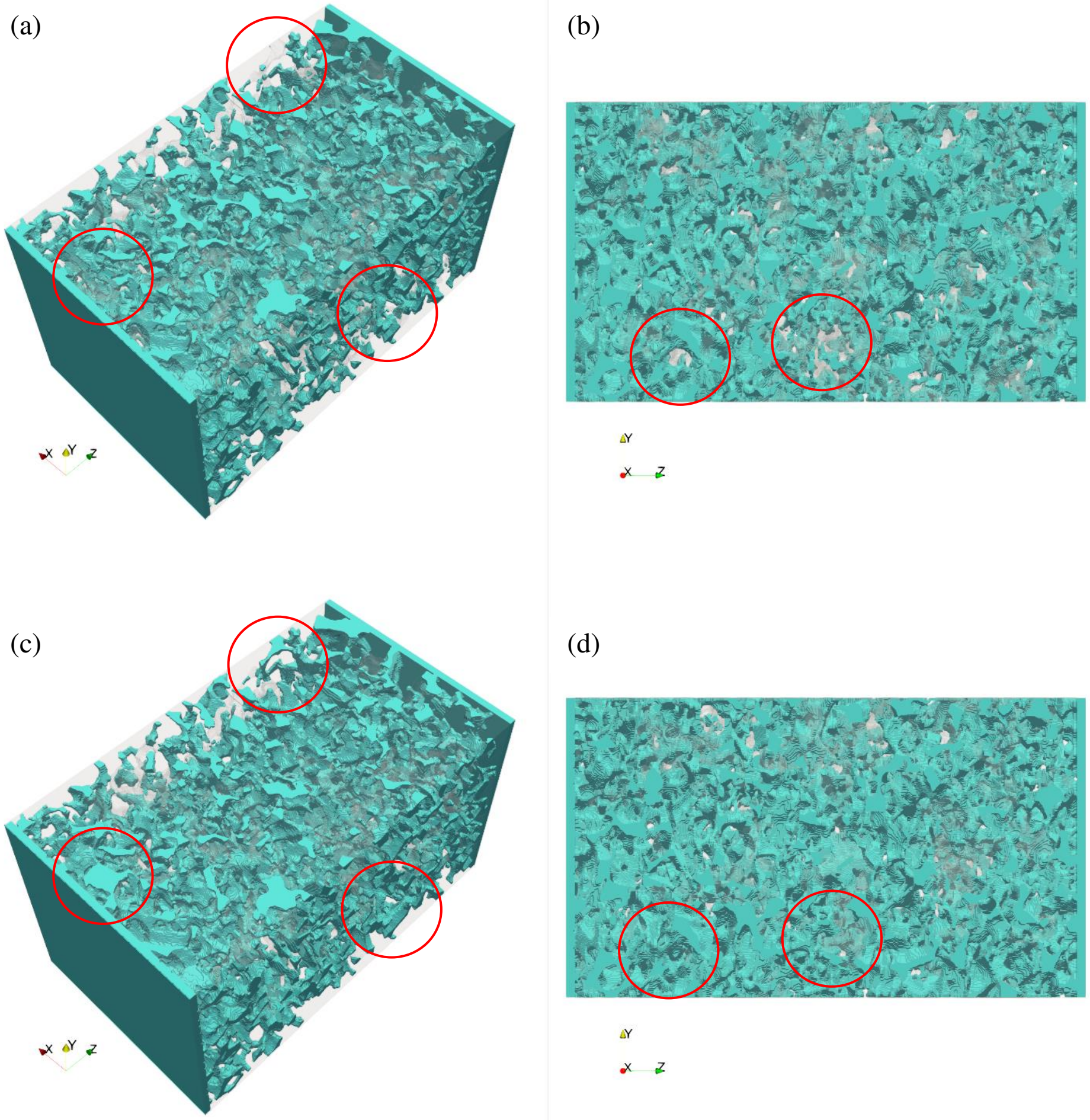

**Figure 16 Comparison between snapshots from the end of spontaneous imbibition simulations. Snapshots (a) and (b) are for Type-I implementation simulations, while snapshots (c) and (d) are for Type-II. Snapshots (a) and (c) show an isometric view of the geometry, while snapshots (b) and (d) show a front view of the geometry. Turquoise color represents water, transparent grey represents solid, and oil is hidden.**

$S_{w,sp} = 0.623$, Type-II: $S_{w,sp} = 0.695$. It could be noted that the difference between the water saturations was further increased in the spontaneous imbibition water saturations. The difference between the spontaneous imbibition simulations can be observed in Figure 16.
For the forced imbibition curves, both types had similar behavior, but Type-II had a slightly steeper curve, which could catch up with the water saturation of Type-I. Both types had a similar final water saturation as follows: Type-I: $S_w = 0.843$, Type-II: $S_w = 0.839$. These values are considered slightly high for the expected final water saturations from laboratory experiments [47]. The low oil saturation could be attributed to the relatively small size of the rock sample, where less oil was trapped during the imbibition displacement process.

## V. Summary

In this work, we investigated the performance of two wettability schemes and two pressure gradient implementations in the lattice Boltzmann color gradient model. We started by validating the CGM model without both the wetting schemes and the pressure gradient implementation using the Laplace test, where it was shown that the model accurately simulated the interfacial tension between two fluids with different viscosities. Then, a static contact angle test was used to assess the accuracy of the wetting models in simulating the desired contact angle of a fluid droplet surrounded by another fluid on a solid surface. It was shown that Scheme-II (proposed by Akai et al. [24]) was more accurate than Scheme-I (proposed by Leclaire et al. [19]).

Then, a dynamic case of a binary fluid flow in a horizontal capillary tube described by the Washburn equation was used to assess the performance of the wetting models and the pressure gradient implementation types. Scheme-II of wetting models was shown to be more accurate than Scheme-I in simulating a spontaneous imbibition displacement process. It was also demonstrated that Type-I of the pressure gradient implementations was more accurate than Type-II in simulating neutrally wetting fluid displacement moving under a pressure gradient.

Finally, two fluid displacement processes, namely the primary drainage and the imbibition, were simulated in a Bentheimer sandstone rock sample. The simulation showed that Type-II pressure gradient implementation resulted in a higher irreducible water (wetting fluid) saturation during the primary drainage displacement. That happened as the oil (non-wetting fluid) could not invade the relatively small pores using this implementation type. Also, it was observed that the water saturation after the spontaneous imbibition process was lower for Type-II. That was related to the different initial water saturation between the two simulations (endpoints of the primary drainage simulations). However, the final water saturation after the forced imbibition displacement process was similar for both implementation types.

## CONFLICTS OF INTEREST

The authors have no conflicts to disclose.

## ACKNOWLEDGMENTS

RCVC acknowledge financial support from the Portuguese Foundation for Science and Technology (FCT) under the contracts: PTDC/FISMAC/5689/2020 (DOI 10.54499/PTDC/FIS-MAC/5689/2020), UIDB/00618/2020 (DOI 10.54499/UIDB/00618/2020), UIDP/00618/2020 (DOI 10.54499/UIDP/00618/2020), DL 57/2016/CP1479/CT0057 (DOI 10.54499/DL57/2016/CP1479/CT0057) and 2023.10412.CPCA.A2.

## APPENDIX A. MRT MODEL DETAILS FOR $D_3Q_{19}$ LATTICE

The $D_3Q_{19}$ lattice arrangement used in this work has the following discrete velocity vectors $\boldsymbol{e}_i$ and lattice weights $w_i$

$$\boldsymbol{e}_i = \begin{bmatrix} 0 & 1 & -1 & 0 & 0 & 0 & 0 & 1 & -1 & 1 & -1 & 1 & -1 & 1 & -1 & 0 & 0 & 0 & 0 \\ 0 & 0 & 0 & 1 & -1 & 0 & 0 & 1 & 1 & -1 & -1 & 0 & 0 & 0 & 0 & 1 & -1 & 1 & -1 \\ 0 & 0 & 0 & 0 & 0 & 1 & -1 & 0 & 0 & 0 & 0 & 1 & 1 & -1 & -1 & 1 & 1 & -1 & -1 \end{bmatrix}, \quad \text{(A.1)}$$

$$w_i = \begin{bmatrix} \frac{1}{3} & \frac{1}{18} & \frac{1}{18} & \frac{1}{18} & \frac{1}{18} & \frac{1}{18} & \frac{1}{18} & \frac{1}{36} & \frac{1}{36} & \frac{1}{36} & \frac{1}{36} & \frac{1}{36} & \frac{1}{36} & \frac{1}{36} & \frac{1}{36} & \frac{1}{36} & \frac{1}{36} & \frac{1}{36} & \frac{1}{36} \end{bmatrix}. \quad \text{(A.2)}$$

The transformation matrix $\boldsymbol{M}$ for the used $D_3Q_{19}$ lattice arrangement is given by

$$\boldsymbol{M} = \begin{pmatrix} 1 & 1 & 1 & 1 & 1 & 1 & 1 & 1 & 1 & 1 & 1 & 1 & 1 & 1 & 1 & 1 & 1 & 1 & 1 \\ -30 & -11 & -11 & -11 & -11 & -11 & -11 & 8 & 8 & 8 & 8 & 8 & 8 & 8 & 8 & 8 & 8 & 8 & 8 \\ 12 & -4 & -4 & -4 & -4 & -4 & -4 & 1 & 1 & 1 & 1 & 1 & 1 & 1 & 1 & 1 & 1 & 1 & 1 \\ 0 & 1 & -1 & 0 & 0 & 0 & 0 & 1 & -1 & 1 & -1 & 1 & -1 & 1 & -1 & 0 & 0 & 0 & 0 \\ 0 & -4 & 4 & 0 & 0 & 0 & 0 & 1 & -1 & 1 & -1 & 1 & -1 & 1 & -1 & 0 & 0 & 0 & 0 \\ 0 & 0 & 0 & 1 & -1 & 0 & 0 & 1 & 1 & -1 & -1 & 0 & 0 & 0 & 0 & 1 & -1 & 1 & -1 \\ 0 & 0 & 0 & -4 & 4 & 0 & 0 & 1 & 1 & -1 & -1 & 0 & 0 & 0 & 0 & 1 & -1 & 1 & -1 \\ 0 & 0 & 0 & 0 & 0 & 1 & -1 & 0 & 0 & 0 & 0 & 1 & 1 & -1 & -1 & 1 & 1 & -1 & -1 \\ 0 & 0 & 0 & 0 & 0 & -4 & 4 & 0 & 0 & 0 & 0 & 1 & 1 & -1 & -1 & 1 & 1 & -1 & -1 \\ 0 & 2 & 2 & -1 & -1 & -1 & -1 & 1 & 1 & 1 & 1 & 1 & 1 & 1 & 1 & -2 & -2 & -2 & -2 \\ 0 & -4 & -4 & 2 & 2 & 2 & 2 & 1 & 1 & 1 & 1 & 1 & 1 & 1 & 1 & -2 & -2 & -2 & -2 \\ 0 & 0 & 0 & 1 & 1 & -1 & -1 & 1 & 1 & 1 & 1 & -1 & -1 & -1 & -1 & 0 & 0 & 0 & 0 \\ 0 & 0 & 0 & -2 & -2 & 2 & 2 & 1 & 1 & 1 & 1 & -1 & -1 & -1 & -1 & 0 & 0 & 0 & 0 \\ 0 & 0 & 0 & 0 & 0 & 0 & 0 & 1 & -1 & -1 & 1 & 0 & 0 & 0 & 0 & 0 & 0 & 0 & 0 \\ 0 & 0 & 0 & 0 & 0 & 0 & 0 & 0 & 0 & 0 & 0 & 0 & 0 & 0 & 0 & 1 & -1 & -1 & 1 \\ 0 & 0 & 0 & 0 & 0 & 0 & 0 & 0 & 0 & 0 & 0 & 1 & -1 & -1 & 1 & 0 & 0 & 0 & 0 \\ 0 & 0 & 0 & 0 & 0 & 0 & 0 & 1 & -1 & 1 & -1 & -1 & 1 & -1 & 1 & 0 & 0 & 0 & 0 \\ 0 & 0 & 0 & 0 & 0 & 0 & 0 & -1 & -1 & 1 & 1 & 0 & 0 & 0 & 0 & 1 & -1 & 1 & -1 \\ 0 & 0 & 0 & 0 & 0 & 0 & 0 & 0 & 0 & 0 & 0 & 1 & 1 & -1 & -1 & -1 & -1 & 1 & 1 \end{pmatrix}. \quad \text{(A.3)}$$

The equilibrium moments $m_i^{eq}$ are given by [8]

$$\begin{aligned} m_0^{eq} &= \rho, \\ m_1^{eq} &= e^{eq} = -11\rho + 19\rho_o\left(u_x^2 + u_y^2 + u_z^2\right), \\ m_2^{eq} &= \epsilon^{eq} = 3\rho - \frac{11}{2}\rho_o\left(u_x^2 + u_y^2 + u_z^2\right), \\ m_3^{eq} &= \rho_o u_x, \\ m_4^{eq} &= -\frac{2}{3}\rho_o u_x, \\ m_5^{eq} &= \rho_o u_y, \\ m_6^{eq} &= -\frac{2}{3}\rho_o u_y, \\ m_7^{eq} &= \rho_o u_z, \\ m_8^{eq} &= -\frac{2}{3}\rho_o u_z, \\ m_9^{eq} &= 3p_{xx}^{eq} = \rho_o\left(2u_x^2 - u_y^2 - u_z^2\right), \\ m_{10}^{eq} &= -\frac{3}{2}p_{xx}^{eq} = -\frac{3}{2}\rho_o\left(2u_x^2 - u_y^2 - u_z^2\right), \\ m_{11}^{eq} &= p_{zz}^{eq} = \rho_o\left(u_y^2 - u_z^2\right), \end{aligned} \quad \text{(A.4)}$$

$$m_{12}^{eq} = -\frac{1}{2} p_{zz}^{eq} = -\frac{1}{2}\rho_o\left(u_y^2 - u_z^2\right),$$
$$m_{13}^{eq} = p_{xy}^{eq} = \rho_o u_x u_y,$$
$$m_{14}^{eq} = p_{yz}^{eq} = \rho_o u_y u_z,$$
$$m_{15}^{eq} = \rho_o u_x u_z,$$
$$m_{16}^{eq} = m_{17}^{eq} = m_{18}^{eq} = 0.$$

The forcing term $\boldsymbol{S^F}$ is expressed as follows

$$\boldsymbol{S^F}(\boldsymbol{x}) = \begin{bmatrix} 0 \\ 38\left(1-\frac{1}{2}S_e\right)\boldsymbol{u}(\boldsymbol{x})\cdot\boldsymbol{F}(\boldsymbol{x}) \\ -11\left(1-\frac{1}{2}S_\epsilon(\boldsymbol{x})\right)\boldsymbol{u}(\boldsymbol{x})\cdot\boldsymbol{F}(\boldsymbol{x}) \\ F_x(\boldsymbol{x}) \\ -\frac{2}{3}\left(1-\frac{1}{2}S_q(\boldsymbol{x})\right)F_x(\boldsymbol{x}) \\ F_y(\boldsymbol{x}) \\ -\frac{2}{3}\left(1-\frac{1}{2}S_q(\boldsymbol{x})\right)F_y(\boldsymbol{x}) \\ F_z(\boldsymbol{x}) \\ -\frac{2}{3}\left(1-\frac{1}{2}S_q(\boldsymbol{x})\right)F_z(\boldsymbol{x}) \\ 2\left(1-\frac{1}{2}S_\nu(\boldsymbol{x})\right)\left(2u_x(\boldsymbol{x})F_x(\boldsymbol{x}) - u_y(\boldsymbol{x})F_y(\boldsymbol{x}) - u_z(\boldsymbol{x})F_z(\boldsymbol{x})\right) \\ \left(1-\frac{1}{2}S_\pi(\boldsymbol{x})\right)\left(-2u_x(\boldsymbol{x})F_x(\boldsymbol{x}) + u_y(\boldsymbol{x})F_y(\boldsymbol{x}) + u_z(\boldsymbol{x})F_z(\boldsymbol{x})\right) \\ 2\left(1-\frac{1}{2}S_\nu(\boldsymbol{x})\right)\left(u_y(\boldsymbol{x})F_y(\boldsymbol{x}) - u_z(\boldsymbol{x})F_z(\boldsymbol{x})\right) \\ \left(1-\frac{1}{2}S_\pi(\boldsymbol{x})\right)\left(-u_y(\boldsymbol{x})F_y(\boldsymbol{x}) + u_z(\boldsymbol{x})F_z(\boldsymbol{x})\right) \\ \left(1-\frac{1}{2}S_\nu(\boldsymbol{x})\right)\left(u_y(\boldsymbol{x})F_x(\boldsymbol{x}) + u_x(\boldsymbol{x})F_y(\boldsymbol{x})\right) \\ \left(1-\frac{1}{2}S_\nu(\boldsymbol{x})\right)\left(u_z(\boldsymbol{x})F_y(\boldsymbol{x}) + u_y(\boldsymbol{x})F_z(\boldsymbol{x})\right) \\ \left(1-\frac{1}{2}S_\nu(\boldsymbol{x})\right)\left(u_x(\boldsymbol{x})F_z(\boldsymbol{x}) + u_z(\boldsymbol{x})F_x(\boldsymbol{x})\right) \\ 0 \\ 0 \\ 0 \end{bmatrix}. \tag{A.5}$$

## APPENDIX B. ZOU-HE PRESSURE BOUNDARY CONDITIONS

Inlet and outlet pressure boundary conditions for the binary fluid CGM model used in this work are implemented using the Zou-He method [26] [43]. However, the technique was originally developed for single-component fluid LBM models. Hence, using the Zou-He technique for a multi-component fluid model requires some treatments. This section summarizes the equations used to implement inlet and outlet pressure boundary conditions for the CGM model.

Inlet pressure boundary conditions could be implemented for the $D_3Q_{19}$ lattice using the Zou-He method, as shown in the following equations. Starting with the fluid component tangential velocity value at the inlet boundary, which is determined by

$$u_z^s(\boldsymbol{x}_{in}) = \rho_{in}\,\zeta^s - [f_0^s + f_1^s + f_2^s + f_3^s + f_4^s + f_7^s + f_8^s + f_9^s + f_{10}^s + 2(f_6^s + f_{13}^s + f_{14}^s + f_{17}^s + f_{18}^s)], \tag{B.1}$$

where, $\zeta^s$ is the fraction of the injected fluid, i.e., $\zeta^r = 1$ and $\zeta^b = 0$ for a pure fluid "r" at the inlet boundary and vice versa. Also, the parameter $\zeta^s$ could be used for a mixed injection of both fluids.

Then, transversal momentum corrections on the z-boundary at the inlet boundary for the distribution functions of each fluid propagating in the x and y-directions ($N_x^s$, $N_y^s$) are determined respectively by

$$N_x^s(\boldsymbol{x}_{in}) = \frac{1}{2}[f_1^s + f_7^s + f_9^s - (f_2^s + f_8^s + f_{10}^s)], \tag{B.2}$$

$$N_y^s(\boldsymbol{x}_{in}) = \frac{1}{2}[f_3^s + f_7^s + f_8^s - (f_4^s + f_9^s + f_{10}^s)], \tag{B.3}$$

Finally, the missing distribution functions of each fluid are determined by

$$f_5^s = f_6^s + \frac{1}{3}[u_z^s(\boldsymbol{x}_{in})], \tag{B.4}$$

$$f_{11}^s = f_{14}^s + \frac{1}{6}[u_z^s(\boldsymbol{x}_{in})] - N_x^s(\boldsymbol{x}_{in}), \tag{B.5}$$

$$f_{12}^s = f_{13}^s + \frac{1}{6}[u_z^s(\boldsymbol{x}_{in})] + N_x^s(\boldsymbol{x}_{in}), \tag{B.6}$$

$$f_{15}^s = f_{18}^s + \frac{1}{6}[u_z^s(\boldsymbol{x}_{in})] - N_y^s(\boldsymbol{x}_{in}), \tag{B.7}$$

$$f_{16}^s = f_{17}^s + \frac{1}{6}[u_z^s(\boldsymbol{x}_{in})] + N_y^s(\boldsymbol{x}_{in}). \tag{B.8}$$

Outlet pressure boundary conditions could be implemented for the $D_3Q_{19}$ lattice using the Zou-He method, as shown in the following equations. Starting with the total tangential velocity value at the outlet boundary, which is determined by

$$u_z^t(\boldsymbol{x}_{out}) = -\rho_{out} + [f_0^t + f_1^t + f_2^t + f_3^t + f_4^t + f_7^t + f_8^t + f_9^t + f_{10}^t + 2(f_5^t + f_{11}^t + f_{12}^t + f_{15}^t + f_{16}^t)]. \tag{B.9}$$

It could be noted that the tangential velocity at the outlet boundary is evaluated differently than the inlet boundary, as shown in Eq. (B.1). At the inlet, a specific fluid (or a mixture of fluids) is injected continuously. Hence, each fluid's tangential velocity is evaluated according to the injection fraction. However, both fluids can leave the domain freely at the outlet. Therefore, a total tangential velocity is used for both fluids. As shown next, the total tangential velocity value will be weighted for each fluid.

Transversal momentum corrections on the z-boundary at the outlet boundary for the distribution functions of each fluid propagating in the x and y-directions are determined respectively by

$$N_x^s(\boldsymbol{x}_{out}) = \frac{1}{2}[f_1^s + f_7^s + f_9^s - (f_2^s + f_8^s + f_{10}^s)], \tag{B.10}$$

$$N_y^s(\boldsymbol{x}_{out}) = \frac{1}{2}[f_3^s + f_7^s + f_8^s - (f_4^s + f_9^s + f_{10}^s)]. \tag{B.11}$$

Finally, the missing distribution functions of each fluid are determined using a weighting function ($\chi^s$) of the phase order parameter $\phi$ as follows

$$\chi^r = \frac{1}{2}\left(1 + \phi(\boldsymbol{x}_{out})\right), \tag{B.12}$$

$$\chi^b = \frac{1}{2}\left(1 - \phi(\boldsymbol{x}_{out})\right), \tag{B.13}$$

$$f_6^s = f_5^s - \frac{1}{3}[u_z^t(\boldsymbol{x}_{out})\chi^s], \tag{B.14}$$

$$f_{13}^s = f_{12}^s - \frac{1}{6}[u_z^t(\boldsymbol{x}_{out})\chi^s] - N_x^s(\boldsymbol{x}_{out}), \tag{B.15}$$

$$f_{14}^s = f_{11}^s - \frac{1}{6}[u_z^t(\boldsymbol{x}_{out})\chi^s] + N_x^s(\boldsymbol{x}_{out}), \tag{B.16}$$

$$f_{17}^s = f_{16}^s - \frac{1}{6}[u_z^t(\boldsymbol{x}_{out})\chi^s] - N_y^s(\boldsymbol{x}_{out}), \tag{B.17}$$

$$f_{18}^s = f_{15}^s - \frac{1}{6}[u_z^t(\boldsymbol{x}_{out})\chi^s] + N_y^s(\boldsymbol{x}_{out}). \tag{B.18}$$